\documentclass[a4paper,11pt]{article}
\pdfoutput=1

\usepackage{jheppub}
\usepackage[T1]{fontenc}

\title{\boldmath Interplay between Higgs inflation and dark matter models with dark $U(1)$ gauge symmetry}

\author[a]{Sarif Khan,}
\author[b,*]{Jinsu Kim,}
\note[*]{Corresponding author.}
\author[c]{and Pyungwon Ko}

\affiliation[a]{
    Institut f\"{u}r Theoretische Physik, 
    Georg-August-Universit\"{a}t G\"{o}ttingen,\\
    Friedrich-Hund-Platz 1,
    37077 G\"{o}ttingen, Germany
}
\affiliation[b]{
    School of Physics Science and Engineering, 
    Tongji University,\\
    Shanghai 200092, China
}
\affiliation[c]{
    Korea Institute for Advanced Study, \\
    Seoul 02455, Republic of Korea
}

\emailAdd{sarif.khan@uni-goettingen.de}
\emailAdd{kimjinsu@tongji.edu.cn}
\emailAdd{pko@kias.re.kr}

\abstract{
We investigate dark matter phenomenology and Higgs inflation in a dark $U(1)_D$-extended model. The model features two dark matter candidates, a dark fermion and a dark vector boson. When the fermion dark matter $\psi$ is heavier than the vector dark matter $W_D$, there is an ample parameter space where $\psi$ is dominant over $W_D$. The model can then easily evade the stringent bounds from direct detection experiments, since $\psi$ has no direct coupling to the Standard Model particles. Furthermore, the model can accommodate inflation in three different ways, one along the Standard Model Higgs direction, one along the dark Higgs direction, and one along the combination of the two. Considering the running of the parameters and various observational constraints, we perform a detailed numerical analysis and identify allowed parameter spaces that explain both dark matter and Higgs inflation in a unified manner. We discuss in detail how the imposition of Higgs inflation severely constrains the dark matter parameter space. The existence of the dark Higgs field is found to play a crucial role both in dark matter phenomenology and in generalised Higgs inflation.
}

\begin{document}
\maketitle
\flushbottom

%%%%%%%%%%%%%%%%%%%%%%%%%%%%%%%%%%%%%%%%%%
\section{Introduction}
\label{sec:intro}
%%%%%%%%%%%%%%%%%%%%%%%%%%%%%%%%%%%%%%%%%%

Identifying physical properties of dark matter (DM) particles such as masses, spins, lifetimes, and any additional quantum numbers is one of the fundamental problems in (astro-)particle physics and cosmology, both for theoretical and experimental physicists.
The most important characteristic of DM would be that DM should either be absolutely stable or have lifetime much longer than the age of the Universe, the latter of which can readily be realised if DM is light enough.

Usually, the DM stability or longevity is associated with some dark symmetry that is either global or local, and exact or approximate. 
In case it is global and approximate, dark symmetry-breaking terms in the Lagrangian should be from small explicit symmetry breaking so that DM lifetime can be much longer than the age of the Universe.  
Another possibility is that a global dark symmetry is an exact one which can protect DM from decaying.\footnote{
This is implicitly assumed in most works on DM physics.
}
However, it is unlikely that a global dark symmetry is exact according to our current understanding of quantum field theory in the presence of gravity.
It is a longtime folklore that any global symmetry would generically be broken in the presence of gravity; see, {\it e.g.}, Ref.~\cite{Harlow:2023pxf} for a recent review.
Notably, there will be $1/M_{\rm Planck}$-scale suppressed dimension-5 operators that violate the global dark symmetry, thereby inducing electroweak (EW)-scale DM to decay too fast to be a good DM candidate.
Assuming that couplings of these dangerous dimension-5 operators are $\sim \mathcal{O}(1)$, one finds that the lifetime of EW-scale DM would become too short \cite{Baek:2013qwa}.\footnote{
For light bosonic (fermionic) DM of mass $m_{\rm boson} \lesssim \mathcal{O}(10)$ keV ($m_{\rm fermion} \lesssim \mathcal{O}(1)$ GeV), its lifetime induced by dimension-5 operators that are suppressed by the Planck scale $M_{\rm Planck}$ can be long enough for DM to be stable in cosmological timescale \cite{Baek:2013qwa}. Invisible axion and light sterile neutrino are good examples.
}
It is a generic problem for the DM stability or longevity relying on a global dark symmetry.
This disastrous situation can be rescued if we consider a local dark gauge symmetry instead of a global dark symmetry to accommodate the DM stability or longevity \cite{Baek:2013qwa,Frigerio:2022kyu}. 
If dark symmetry is an exact one, and the DM particle is absolutely stable, it is better to implement a dark local gauge symmetry, just like in QCD or QED in the Standard Model (SM). 
In this way, the dark gauge boson (or dark photon) enters the game in a natural manner.

Amongst possible dark gauge symmetry groups, Abelian $U(1)_D$ is the simplest and has been most studied in various contexts; see Ref.~\cite{Fabbrichesi:2020wbt} for a recent review and references therein.
DM models with a dark photon belong to this category, and the kinetic mixing between the dark $U(1)_D$ and the SM $U(1)_Y$ or $U(1)_{\rm em}$ field-strengths is generically allowed at the renormalisable level.  
In case the dark photon is massless, DM is mini-charged after the kinetic mixing is removed by the field redefinition \cite{Fabbrichesi:2020wbt}, and one can envisage interesting phenomenology in both (astro-)particle physics and cosmology; see Refs. \cite{Kaplan:2009de,McDermott:2010pa,Munoz:2018pzp} for some cases of the massless dark photon and mini-charged DM.
If the dark photon is massive, on the other hand, the SM fields can couple to the dark photon through the $U(1)$ kinetic mixing \cite{Fabbrichesi:2020wbt}. If one assumes that the kinetic mixing is very small or forbidden by dark $CP$ symmetry, massive dark photon can make a vector DM. The massive dark photon case is usually described in the St\"{u}ckelberg mechanism, without paying much attention to the origin of dark photon mass.
However, this approach sometimes suffers from unitarity violation in the high energy or massless dark photon limit, and physics results obtained within that framework could be misleading, catastrophic, or puzzling.
This issue could be resolved when one includes an agency that provides the dark photon mass. The simplest well-known way is to consider dark Higgs mechanism where the dark Higgs is also charged under the dark gauge group and develops a nonzero vacuum expectation value (VEV).
Once the dark Higgs comes into play, unitarity will be restored and some puzzles for the massless dark photon limit in the Higgs decay width into the pair of dark photons  disappear~\cite{Baek:2014jga,Baek:2021hnl}.
Also, we can observe how the unitarity is violated/restored in models without/with the dark Higgs field in the DM pair productions at the International Linear Collider~\cite{Ko:2016xwd}.
Finally, the dark Higgs field opens new channels for DM pair annihilations in the $p$-wave, and one can realise the light thermal WIMP scenarios evading most stringent bounds from CMB constraints \cite{Baek:2020owl}.
Depending on the charge assignments of DM and dark Higgs fields, one can enjoy very rich phenomenology with theoretically and mathematically consistent frameworks.
The dark $U(1)_D$ can be completely broken, which could be a generic case, in which the situation becomes similar to the global dark symmetry cases with explicit symmetry breaking. 
The dark $U(1)_D$ can also be broken to its $Z_N$ subgroups; for general discussions, see Refs.~\cite{Batell:2010bp,Yaguna:2019cvp}, and specific discussions on some detailed DM phenomenology including the importance of the dark Higgs boson can be found in, {\it e.g.}, Refs.~\cite{Baek:2014kna,Ko:2019wxq,Baek:2020owl} for $Z_2$ scalar or fermion DM, and Refs.~\cite{Ko:2014nha,Ko:2014loa} for $Z_3$ complex scalar DM.

For dark gauge symmetries, one can also consider non-Abelian dark gauge symmetries, where DM model buildings and phenomenology become even more interesting and richer.
For example, one can consider a possibility that the dark gauge sector is confining, similar to the QCD sector in the SM \cite{Hur:2007uz,Bai:2010qg,Hur:2011sv,Hatanaka:2016rek,Ko:2017uyb}.
In this case, there can appear a new mass scale without a Higgs field due to the dimensional transmutation in the strongly interacting, confining dark sector, and the lightest dark mesons and baryons could make good DM candidates.
Furthermore, both the EW symmetry breaking and cold DM can originate from the strongly interacting hidden sector \cite{Hur:2007uz,Hur:2011sv}. 
Alternatively, one may consider a perturbative dark $SU(N)$, which is broken completely \cite{Boehm:2014bia,Gross:2015cwa} or to its subgroup such as the continuous $SU(M)$ (with $M<N$) \cite{Ko:2016fcd} or discrete $Z_N$ subgroups \cite{Adulpravitchai:2011ei} including $N$-ality \cite{Frigerio:2022kyu}.
If a dark $SU(2)$ is broken to $U(1)$ by a VEV of a real triplet dark Higgs field, there could also appear a topological soliton, such as the dark monopole that can make another stable DM for topological reasons~\cite{Baek:2013dwa}.
In this work, however, we shall consider a simple dark $U(1)_D$ gauge symmetry and study the interplay between DM phenomenology and Higgs inflation.

For dark Higgs fields, one may consider two types: a pure gauge singlet $S$ or a $\phi_D$ with nonzero dark charge. A singlet $S$ has nothing to do with the dark photon mass, since it does not carry any dark charge.\footnote{
We call $S$ a dark Higgs, although it does not carry any dark charge, for convenience.
} 
The singlet $S$ can have renormalisable and gauge-invariant couplings to the SM sector through the $H^\dagger H$ operator, where $H$ is the SM Higgs field, as well as to the dark sector, such as to $\overline{\psi} \psi$ for a dark fermion $\psi$ or $\phi^\dagger \phi$ for a dark scalar $\phi (\neq \phi_D)$.\footnote{
In principle, dark sector fields (DM, dark Higgs, {\it etc.}) can carry nonzero SM quantum numbers, and it would be straightforward to extend the above discussions to such cases.
}
For a dark Higgs $\phi_D$ with a nonzero dark charge, one can consider suitable matter contents and their dark charges such that composite operators made of them have renormalisable and gauge-invariant couplings to $\phi_D$. Depending on the VEV of the dark Higgs $\phi_D$, one can have a massless or massive dark gauge boson.
This way, not only DM particles but also dark gauge bosons (or dark photons) and dark Higgs fields become key players in the dark sector in a natural manner; for a review, see Refs.~\cite{Ko:2016yfb,Ko:2018qxz}.

Another important problem in cosmology is to embed cosmic inflation in underlying particle physics models that describe microscopic world. Within the SM, there is only one scalar field that can play the role of inflaton which drives inflation: the SM Higgs field. This SM Higgs inflation model~\cite{Cervantes-Cota:1995ehs,Bezrukov:2007ep}, with the help of the so-called nonminimal coupling of the SM Higgs field to gravity of the form $|H|^2 R$, where $R$ is the Ricci scalar, is one of the most favoured models by the latest observations~\cite{Planck:2018jri,BICEP:2021xfz}; see, {\it e.g.}, Refs.~\cite{Cheong:2021vdb,Rubio:2018ogq} for a recent review. The presence of such a nonminimal coupling term may be seen natural as it has mass-dimension of four. Furthermore, the term would generically arise through radiative corrections; see, {\it e.g.}, Ref.~\cite{Buchbinder:1992rb}.
The SM Higgs inflation model requires the nonminimal coupling parameter to be $\simeq 47000\sqrt{\lambda_H}$, where $\lambda_H$ is the SM Higgs quartic coupling, to match the amplitude of the curvature power spectrum. For $\lambda_H \simeq 0.125$, for instance, the nonminimal coupling parameter should thus be large $\sim 10^4$.\footnote{
Once the renormalisation group (RG) running is taken into account, it is possible to realise a scenario where $\lambda_H$ becomes tiny at the inflation scale in which case the nonminimal coupling parameter could take a small value. While such a scenario is hard to realise in the pure SM as the top-quark pole mass needs to be about $3\sigma$ away from the central value \cite{Buttazzo:2013uya}, extensions of the SM could make such a scenario possible.
}
Such a large nonminimal coupling parameter has led to the discussion of the unitarity violation \cite{Burgess:2009ea,Barbon:2009ya,Lerner:2009na,Burgess:2010zq,Hertzberg:2010dc}. In Ref.~\cite{Bezrukov:2010jz}, it is shown that the cutoff scale coming from the unitarity violation depends on the background inflaton field value, and the cutoff scale becomes larger than the relevant scale of inflation.
In DM models with local dark gauge symmetries, there would appear additional scalar fields, namely dark Higgs fields, which can also play the role of inflaton, inducing dark Higgs inflation just like the SM Higgs inflation. The dark Higgs fields may also nonminimally couple to gravity. For instance, a singlet $S$ can have nonminimal couplings of the form $S R$ and/or $S^2 R$, while a dark Higgs field with nonzero dark charge can have $|\phi_D|^2 R$.
We note that the quartic couplings of the dark Higgs fields could be small. As such, the corresponding nonminimal coupling parameters need not be as large as the one required in the pure SM Higgs inflation model.

Connecting between cosmic inflation and DM physics in models with extra symmetries is an interesting subject; see, {\it e.g.}, Refs.~\cite{Lerner:2009xg,Gong:2012ri,Haba:2014zda,Kim:2014kok,Aravind:2015xst,Choudhury:2015eua,Tenkanen:2016idg,Ballesteros:2016xej,Hooper:2018buz,Choi:2019osi,Choi:2020ara,Kawai:2021tzc,Ghoshal:2022jeo,Cheng:2022hcm,Qi:2023egb,Kawai:2023vjo}. In order to link two different scales, it is vital to connect coupling parameters at different energy scales with the RG running. In the pure SM case, once quantum corrections are taken into account, the quartic coupling of the SM Higgs field $\lambda_H$ falls below zero as we move to high-energy scales. As such, the SM Higgs inflation may become unstable. On the other hand, in DM models with local dark gauge symmetries, the existence of extra dark Higgs fields may lift the SM Higgs quartic coupling in such a way that it stays positive all the way up to the inflation scale, resurrecting the SM Higgs inflation \cite{Kim:2014kok}. Moreover, as there are more than one scalar field, inflation may occur along the SM Higgs direction, the dark Higgs direction, or the combination of the two.
In this work, we will study correlations between DM phenomenology and Higgs inflation in a dark $U(1)_D$-extended model. We shall perform both the classical and quantum analyses and map different inflation scenarios in a parameter space. 

This paper is organised as follows. In Sec.~\ref{sec:model}, we propose a model based on the dark $U(1)_D$, which is supposed to be broken by the nonzero VEV of dark Higgs field $\phi_D$ with dark charge $1$. We also introduce dark fermion $\psi$ with $U(1)_D$ charge equal to $n_\psi$, which will be (a part of) DM of the Universe. Section~\ref{sec:DM} discusses DM phenomenology in detail. In particular, we shall focus on a two-component DM scenario, where dominant component of DM is almost hidden from direct detection experiments. In Sec.~\ref{sec:Inf}, we discuss Higgs inflation with or without dark Higgs field, both at classical level and at quantum level. In Sec.~\ref{sec:DMInf}, correlations between DM physics and inflation, focusing on the SM Higgs inflation scenario, are discussed. We then summarise the paper in Sec.~\ref{sec:conc}. Explicit expressions for various scattering cross-sections and RG equations relevant to our study are presented in Appendices \ref{apdx:XSecs} and \ref{apdx:RGEs}, respectively.

%%%%%%%%%%%%%%%%%%%%%%%%%%%%%%%%%%%%%%%%%%
\section{Model}
\label{sec:model}
%%%%%%%%%%%%%%%%%%%%%%%%%%%%%%%%%%%%%%%%%%
We consider\footnote{
We use the metric convention of $(-,+,+,+)$.
}
\begin{align}
    \mathcal{L} &= 
    \mathcal{L}_{\overline{\rm SM}}
    +\bar{\psi}\left(
    i\gamma^{\mu}D_{\mu} - M_\psi 
    \right)\psi
    -|D\phi_D|^2
    -|D H|^2
    -V(\phi_D,H)
    \,,
\end{align}
where $\mathcal{L}_{\overline{\rm SM}}$ is the SM Lagrangian barring the SM Higgs sector, $H$ ($\phi_D$) is the SM (dark) Higgs field, the covariant derivative for the dark sector is defined as $D_{\mu}A = \partial_{\mu} A - i g_{D} W_{D} n_{A} A$, with $A=\{\psi,\phi_D\}$, $g_{D}$ is the dark $U(1)_D$ gauge coupling, $W_{D}$ is the gauge boson, usually called the dark photon, associated with $U(1)_D$, and $n_{A}$ is the $U(1)_D$ charge. 
We choose $n_{\phi_D} = 1$ without loss of generality and consider $n_\psi \geq 1$; the $n_\psi = 1/2$ case is thoroughly studied in Refs.~\cite{Baek:2014kna,Baek:2020owl} and Refs.~\cite{Ahmed:2017dbb,Baek:2020owl} for scalar and fermion DM, respectively. Pure vector DM with dark Higgs mechanism was proposed in Refs.~\cite{Farzan:2012hh,Baek:2012se}, and the comparison with the effective field theory is discussed in Refs.~\cite{Baek:2014jga,Baek:2021hnl} in detail.

\begin{center}
\begin{table}[t!]
\begin{center}
\begin{tabular}{||c|c|c|c||}
\hline
\hline
\begin{tabular}{c}
    Gauge\\
    Group\\ 
    \hline
    ${\rm SU(2)}_{\rm L}$\\  
    \hline
    ${\rm U(1)}_{\rm Y}$\\  
\end{tabular}
&
\begin{tabular}{c|c|c}
    \multicolumn{3}{c}{Baryon Fields}\\ 
    \hline
    $Q_{L}^{i}=(u_{L}^{i},d_{L}^{i})^{T}$&$u_{R}^{i}$&$d_{R}^{i}$\\ 
    \hline
    $2$&$1$&$1$\\ 
    \hline
    $1/6$&$2/3$&$-1/3$\\  
\end{tabular}
&
\begin{tabular}{c|c}
    \multicolumn{2}{c}{Lepton Fields}\\
    \hline
    $L_{L}^{i}=(\nu_{L}^{i},e_{L}^{i})^{T}$ & $e_{R}^{i}$\\
    \hline
    $2$&$1$\\
    \hline
    $-1/2$&$-1$\\
\end{tabular}
&
\begin{tabular}{c}
    \multicolumn{1}{c}{Scalar Field}\\
    \hline
    $H$\\
    \hline
    $2$\\
    \hline
    $1/2$\\
\end{tabular}\\
\hline
\hline
\end{tabular}
\caption{SM particle contents and their corresponding charges under the SM gauge groups.}
\label{tab:modelSM}
\end{center}
\end{table}
\end{center}
\begin{center}
\begin{table}[t!]
\begin{center}
\begin{tabular}{||c|c|c||}
\hline
\hline
\begin{tabular}{c}
    Gauge\\
    Group\\ 
    \hline
    $U(1)_{D}$\\ 
\end{tabular}
&
\begin{tabular}{c}
    \multicolumn{1}{c}{Fermionic Fields}\\ 
    \hline
    $\psi$\\ 
    \hline
    $n_\psi$\\ 
\end{tabular}
&
\begin{tabular}{c}
    \multicolumn{1}{c}{Scalar Field}\\
    \hline
    $\phi_D$\\
    \hline
    $1$\\
\end{tabular}\\
\hline
\hline
\end{tabular}
\caption{Dark particle contents and their corresponding charges under the additional Abelian $U(1)_D$ gauge group. They are all SM singlets.}
\label{tab:modelU1D}
\end{center}
\end{table}
\end{center}

The scalar potential $V$ is given by
\begin{align}
    V(\phi_D,H) = 
    - \mu^2_D \phi^{\dagger}_D \phi_D
    + \lambda_D (\phi^{\dagger}_D \phi_D)^2 
    - \mu^2_H H^{\dagger} H
    + \lambda_H (H^{\dagger} H)^2 
    + \lambda_{HD} \phi^{\dagger}_D \phi_D H^{\dagger} H
    \,.\label{eqn:scalar_pot_full}
\end{align}
In unitary gauge, one may express the SM and dark Higgs fields as
\begin{align}
    H =
    \frac{1}{\sqrt{2}}
    \begin{pmatrix}
    0 \\
    v_H + h  
    \end{pmatrix}
    \,,\quad
    \phi_D =
    \frac{v_D + \phi}{\sqrt{2}}
    \,,
\end{align}
where $v_H$ and $v_D$ are the VEVs of the SM Higgs field and the dark Higgs field, respectively. The mass matrix for the scalars in the basis of $(h, \phi)$ is given by
\begin{align}
    M_{h\phi} = \begin{pmatrix}
    2\lambda_H v_H^2 & \lambda_{HD} v_H v_D\\
    \lambda_{HD} v_H v_D & 2\lambda_D v_D^2
    \end{pmatrix}
    \,,
\end{align}
while the mass eigenstates are as follows:
\begin{align}
    \begin{pmatrix}
    h_{1}\\
    h_{2}
    \end{pmatrix}
    = \begin{pmatrix}
    \cos\theta & -\sin\theta\\
    \sin\theta & \cos\theta
    \end{pmatrix}
    \begin{pmatrix}
    h\\
    \phi
    \end{pmatrix}\,.
\end{align}
We summarise the SM and dark sector particle contents as well as their corresponding charges in Tables~\ref{tab:modelSM} and \ref{tab:modelU1D}.

In general, the dark $U(1)_D$ gauge boson can couple to the $U(1)_Y$ gauge boson through the gauge kinetic mixing term, through which it will decay into the SM particles. 
In this work, we assume the charge conjugation invariance in the dark sector \cite{Ahmed:2017dbb}:
\begin{align}
\phi_D \rightarrow \phi_D^\dagger
\,,\quad
W_{D\mu} \rightarrow - W_{D\mu}
\,,\quad
\psi \rightarrow \psi^C \equiv - i \gamma_2 \psi^*
\,,
\end{align}
which forbids the $U(1)$ kinetic mixing term.\footnote{
One may alternatively consider a tiny kinetic mixing angle so that the dark $U(1)_D$ gauge boson lives longer than the age of the Universe. In this case, the kinetic mixing parameter needs to be smaller than $\mathcal{O}(10^{-26})$ \cite{Costa:2022lpy}.
}
Once the $U(1)_{D}$ symmetry gets broken, the additional gauge boson acquires the mass of
\begin{align}
    M_{W_{D}} = g_D v_D
    \,.
\end{align}
%%

%%%%%%%%%%%%%%%%%%%%%%%%%%%%%%%%%%%%%%%%%%
\section{Dark matter phenomenology}
\label{sec:DM}
%%%%%%%%%%%%%%%%%%%%%%%%%%%%%%%%%%%%%%%%%%

%%
\begin{figure}[t!]
    \centering
    \includegraphics[scale=0.8]{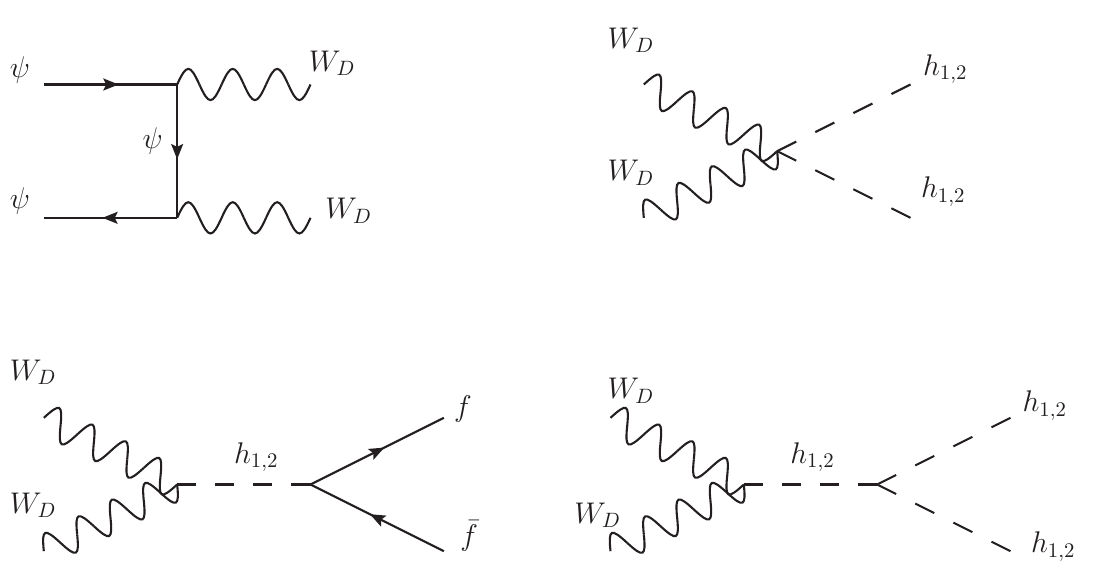}
    \caption{Feynman diagrams relevant for our DM analysis.}
    \label{fig:feynman-diag}
\end{figure}

The model under consideration contains two DM candidates; one is the gauge boson $W_{D}$ associated with the dark $U(1)_D$, and the other is the dark fermion $\psi$. 
The Boltzmann equations for the yields, $Y_i$ ($i=\{\psi,W_D\}$), are given by
\begin{align}
    \frac{d Y_i}{d x} = - \frac{M_{W_{D}}}{x^{2}} \frac{1}{3 H(T)} 
    \frac{d s}{d T} \langle \sigma v \rangle_{ii} \left( Y_i^2 - Y^{{\rm eq},2}_i \right)\,, 
    \label{eqn:BE}
\end{align}
where $x=M_{W_D}/T$, $H(T) = \sqrt{\pi^2 g_{\rho}(T)/90}(T^{2}/M_{\rm P})$ is the Hubble parameter, $s(T) = (2 \pi^{2}/45) g_{s}(T) T^{3}$ is the entropy density, $g_{s}(T)$ and $g_{\rho}(T)$ are the entropic and matter degrees of freedom of the Universe, and $M_{\rm P}$ is the reduced Planck mass. 
Figure~\ref{fig:feynman-diag} shows the Feynman diagrams relevant for our DM study. The relevant cross-sections are given in Appendix~\ref{apdx:XSecs}.
Once the yield is given, the DM relic density can be determined as
\begin{align}
    \Omega_i h^2 = 2.755 \times 10^{8} \left( \frac{M_i}{\rm GeV} \right) Y_i\,.
    \label{eqn:relic-density-expression}
\end{align}
To determine the DM relic density, we have used \texttt{micrOMEGAs}~\cite{Belanger:2001fz}, which essentially solves the Boltzmann equations mentioned above, together with \texttt{FeynRules}~\cite{Alloul:2013bka} and \texttt{CalcHEP}~\cite{Belyaev:2012qa}.

In our analysis of DM phenomenology, we have considered the following constraints, derived from various terrestrial to space-based experiments:

\begin{figure}[t!]
    \centering
    \includegraphics[scale=0.9]{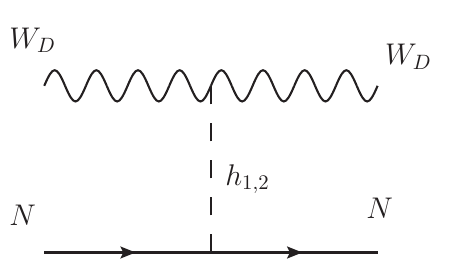}
    \includegraphics[scale=0.6]{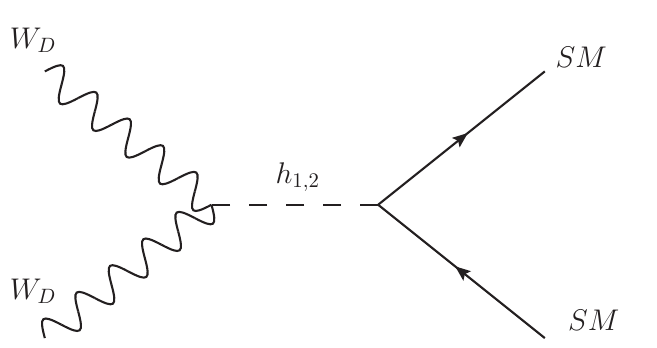}
    \caption{Relevant Feynman diagrams for the direct detection (left) and indirect detection (right) prospects.}
    \label{fig:DDID}
\end{figure}
\begin{itemize}
\item {\bf DM relic density:} 
The total DM relic density, $\Omega_{\rm DM}h^2$, is given by the sum of the relic densities of each DM component, $\Omega_{W_D}h^2$ and $\Omega_{\psi}h^2$.
We have taken the upper range of the DM relic density put by the Planck data~\cite{Planck:2015fie,Planck:2018vyg} and chosen the lower value of DM relic density to be $10^{-4}$,
\begin{align}
    10^{-4} \leq 
    \Omega_{\rm DM}h^{2} \,
    (=\Omega_{W_{D}}h^{2} + \Omega_{\psi}h^{2}) 
    \leq 0.1226 
    \,.\label{eqn:DM-RD-3sigma}
\end{align}
The lower value of DM total relic density $10^{-4}$ is deliberately chosen to boost the computational runtime.
The $\psi$ DM relic density depends on its dark $U(1)_D$ charge as $\Omega_\psi h^2 \propto n_\psi^{-4}$. It indicates that a parameter set that gives rise to a low DM relic density can readily be adjusted to achieve the desired DM relic density by reducing the $U(1)_{D}$ charge of the dark fermion $n_\psi$. 
We note that the shift in $n_{\psi}$ has no effect on the $W_{D}$ contribution to the total DM relic density. Moreover, due to the non-dependence on $n_\psi$, the direct and indirect detection of the $W_{D}$ DM are not affected.
The reduction of $n_\psi$ has thus no adverse impact on other aspects of our work.

\item {\bf Collider bounds:}
The additional SM-neutral Higgs can dominantly decay to $W^{+}W^{-}$, $ZZ$, $f_{\rm SM} \overline{f}_{\rm SM}$, and $W_{D}W_{D}$. 
Amongst the three modes, the first two decay modes further decay to SM particles, while the last mode becomes missing energy. On the other hand, the SM Higgs can decay to the DM sector as well and may contribute as missing energy at the collider. In particular, the interference between the SM and dark Higgs bosons can be important in certain parameter space for both fermion and vector DM \cite{Baek:2011aa,Baek:2015lna,Ko:2016xwd,Ko:2016ybp,Kamon:2017yfx,Dutta:2017sod,Ko:2018mew}.
Moreover, there is a precise measurement of Higgs signal strength which can further constrain the Higgs mixing angle $\theta$ (see, for example, Ref.~\cite{Cheung:2015dta}). In order to consider all of these bounds, we have used \texttt{HiggsBounds}~\cite{Bechtle:2015pma}, which mainly constrains the beyond-the-SM Higgs, and \texttt{HiggsSignal}~\cite{Bechtle:2013xfa}, which mainly constrains the SM Higgs. All the data points presented in the resultant plots have passed those checks.

\item {\bf Direct detection:}
We note that, while $\psi$ has no direct detection prospects, $W_{D}$ may be detected by the WIMP-type DM direct detection experiments, as shown in the left panel (LP) of Fig.~\ref{fig:DDID}.
The analytical estimate for $W_{D} N \rightarrow W_{D} N$ ($N$ is nucleon) takes the form~\cite{Baek:2012se},
\begin{align}
    \sigma_{\rm SI} = \frac{\mu_*^{2}\,\sin^{2}2\theta\,g^2_{D}}{4 \pi v_h^{2}} 
    \left(\frac{1}{M^2_{h_1}} - \frac{1}{M^2_{h_2}} \right)^{2} 
    \left[ \frac{Z \tilde{f}_{p} + (A-Z)\tilde{f}_{n}}{A} \right]^{2}
    \,,\label{eqn:DD-expression}
\end{align}
where $\mu_* = M_{W_D} M_{N}/(M_{W_D} + M_{N})$ is the reduced mass, $M_{N}$ is the nucleon mass, $Z$ ($A$) is the atomic (mass) number, and $\tilde{f}_{\alpha}$ $(\alpha = p,n)$ can be expressed as
\begin{align}
    \frac{\tilde{f}_{\alpha}}{M_{N}} = \left( \frac{7}{9} \sum_{q=u,d,s} 
    f^{\alpha}_{T_q} + \frac{2}{9}\right)\,,
\end{align}
with $f^{p(n)}_{T_u} = 0.020 (0.026)$, $f^{p(n)}_{T_d} = 0.026 (0.020)$, and $f^{p,n}_{T_s} = 0.043$~\cite{Junnarkar:2013ac}. We shall show that, using the spin-independent direct detection cross-sections, a portion of the parameter space could already be ruled out by the LUX-ZEPLIN data~\cite{LZ:2022lsv}.

\item {\bf Indirect Detection:}
The DM candidate $W_D$ may also annihilate to SM particles and can be detected at indirect detection experiments. The generic process by which the DM can be detected is shown in the right panel (RP) of Fig.~\ref{fig:DDID}. The thermal average of cross-section times velocity for the process, $W_{D}W_{D} \rightarrow A A$, with $A$ being SM particles, can be expressed as
\begin{align}
    \langle \sigma v \rangle_{W_{D}W_{D} \rightarrow AA} =  
    \frac{1}{8 M_{W_{D}}^{4} 
    K^2_{2}\left(M_{W_{D}}/T\right) } 
    \int_{4 M^2_{W_{D}}}^{\infty} d s \, 
    \frac{\sigma_{W_{D}W_{D} \rightarrow AA}}{\sqrt{s}} p_{W_{D}} 
    K_{1}\left(\frac{\sqrt{s}}{T}\right) 
    \,,
    \label{eqn:thermal-average-define}
\end{align}
where $p_{W_{D}} = s (s - 4 M^2_{W_{D}})$ and $K_{i}(x)$ is the modified Bessel function of the second kind for the $i^{th}$ order. Expressions for the relevant cross-sections are given in Appendix~\ref{apdx:XSecs}. 
In the resultant plots, we shall show indirect detection bounds associated with $b\bar{b}$ and $W^{+}W^{-}$ channels.
\end{itemize}

For the numerical analysis, we have varied the input model parameters as follows:
\begin{gather}
    50 \leq M_{h_2}\,[{\rm GeV}] \leq 1050 
    \,,\quad
    50 \leq M_{W_D}\,[{\rm GeV}] \leq 1050
    \,,\quad
    1 \leq \left(M_{\psi} - M_{W_{D}}\right)\,\,[{\rm GeV}] \leq 100
    \,,\nonumber\\
    10^{-3} \leq g_{D} \leq 1
    \,,\quad
    10^{-3} \leq \sin\theta \leq 0.5
    \,,\quad
    1 \leq n_{\psi} \leq 100
    \,.
    \label{eqn:random-scanning}
\end{gather}
We have considered $M_{\psi} > M_{W_{D}}$ so that we always have $\psi\bar{\psi} \rightarrow W_{D}W_{D}$ annihilation mode open and assist $\psi$ DM to freeze out when its annihilation rate is smaller than the Hubble rate. Moreover, we have varied $n_\psi \geq 1$. The $n_{\psi} < 1$ scenario will make the $\psi$ DM departure from the thermal bath earlier which results in overproducing the $\psi$ DM for most of the $n_{\psi}$ values. Therefore, to be on the safe side from the Planck upper bound on the DM relic density, we have focused on $n_{\psi} \geq 1$ so that for most of the $n_{\psi}$ values, we do not overproduce $\psi$ DM candidate.
After varying the parameters, we have selected points which satisfy the DM relic density constraint \eqref{eqn:DM-RD-3sigma} and the perturbativity constraint, in particular, $n_\psi g_D < \sqrt{4\pi}$.
In the following, we show resultant plots which exhibit correlations amongst the model parameters. We also discuss various DM observables such as the DM relic density, direct detection cross-section, and indirect detection cross-section. 

\begin{figure}[t!]
    \centering
    \includegraphics[angle=0,height=7.5cm,width=7.5cm]{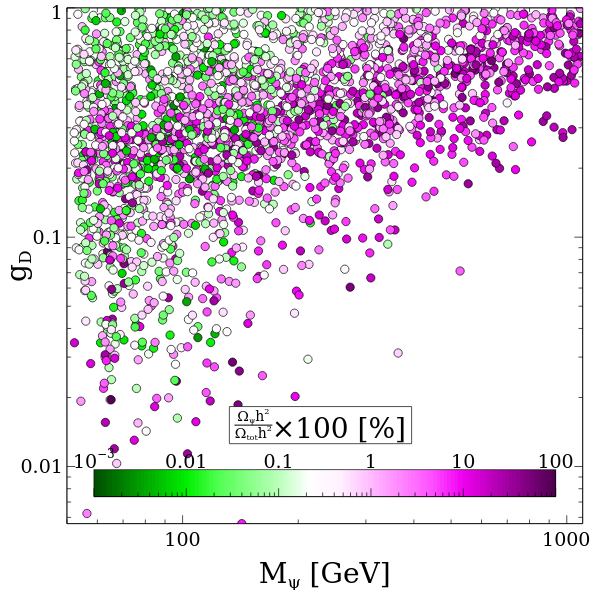}
    \includegraphics[angle=0,height=7.5cm,width=7.5cm]{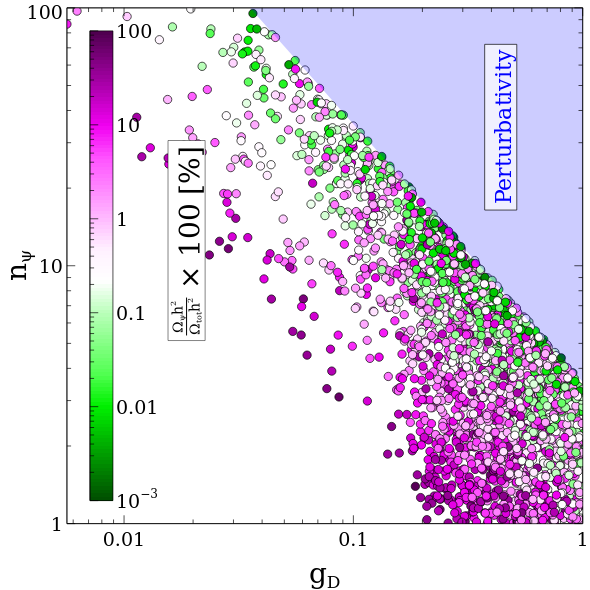}
    \caption{Ratio between the relic density of $\psi$ and the total DM relic density in the $M_{\psi}$--$g_{D}$ plane (left) and $g_{D}$--$n_{\psi}$ plane (right). All the points satisfy the relic constraint~\eqref{eqn:DM-RD-3sigma}. The colour bar represents the percentage of the $\psi$ contribution in total DM relic density. The blue-shaded region corresponds to $n_{\psi} g_{D} \geq \sqrt{4 \pi}$ which violates the perturbativity.} 
    \label{fig:scatter-plot-0}
\end{figure}

Figure~\ref{fig:scatter-plot-0} shows the $\psi$ contribution to the total DM relic density in the $M_{\psi}$--$g_{D}$ plane (LP) and $g_{D}$--$n_{\psi}$ plane (RP), with the colour representing the percentage of the $\psi$ relic in the total DM relic density.
The only process which governs the DM relic density for $\psi$ is shown in Fig.~\ref{fig:feynman-diag}, from which it is clear that the cross-section of the process would depend on the gauge coupling $g_{D}$, the $U(1)_D$-charge $n_{\psi}$, and the mass gap between the initial and final particles. Moreover, the DM relic density is also proportional to the DM mass as shown in Eq.~\eqref{eqn:relic-density-expression}. 
From the LP, we see that the region where the $\psi$ contribution is negligible situates in the top-left corner; it is mainly due to the fact that higher values of $g_D$ lead to more efficient annihilation. The same region also contains cases where the $\psi$ contribution is dominant; it is mainly due to a close mass gap between the initial and final particles and/or smaller values of $n_{\psi}$. 
As $M_{\psi}$ increases, we mainly see the $\psi$-dominant cases because of the linear dependence of the DM density on its mass.
For smaller values of $g_{D} < 0.1$, we get $\geq 10 \%$ of the $\psi$-contribution in the total DM density.
From the RP of Fig.~\ref{fig:scatter-plot-0}, we see that the top-right corner consists of negligible $\psi$-contributions; it is mainly due to larger values of both $n_{\psi}$ and $g_{D}$. 
The blue-shaded region represents $n_{\psi} g_{D} \geq \sqrt{4 \pi}$, {\it i.e.}, violation of the perturbativity. We note that data points that violate the perturbativity are less important due to the negligible contribution to the DM relic density. We observe anti-correlation between the magenta points which represent the $\psi$-dominant cases; this happens as we get $\geq 50 \%$ contributions for particular values of the product $n_{\psi} g_{D}$. 
We stress that Fig.~\ref{fig:scatter-plot-0} serves to showcase the dependence behaviour of the DM relic density on the parameters.
As we discussed earlier, a data point that results in a low total DM relic density can be adjusted to obtain the desired DM relic density by shifting $n_\psi$.
The reduction of $n_\psi$ shall push, for instance, green points, which represent the case of smaller fraction of dark fermion in the total DM relic density, towards magenta points, which represent the case of larger fraction of dark fermion DM relic. Importantly, except when we explicitly exhibit the fraction of dark fermion contribution to the total DM relic density, other plots will remain unchanged by the shift of $n_\psi$, maintaining the final conclusion intact.

\begin{figure}[t!]
    \centering
    \includegraphics[angle=0,height=7.5cm,width=7.5cm]{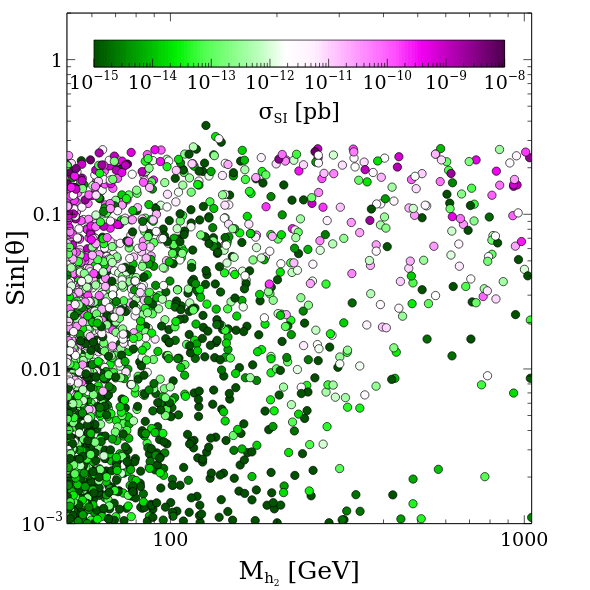}
    \includegraphics[angle=0,height=7.5cm,width=7.5cm]{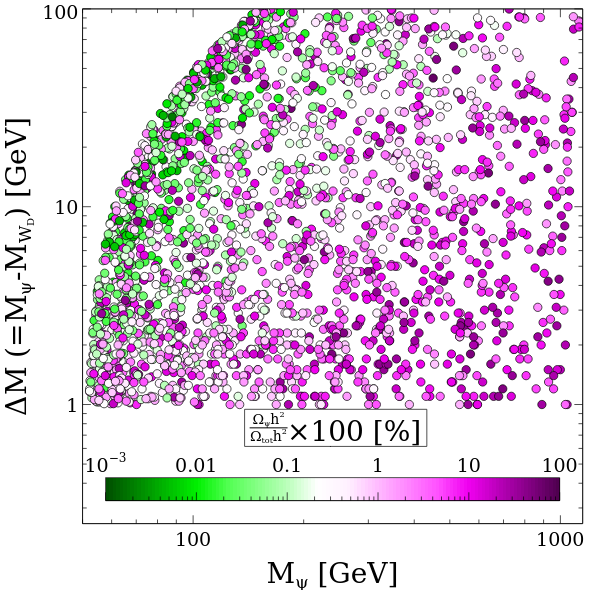}
    \caption{
    Spin-independent cross-section in the $M_{h_{2}}$--$\sin\theta$ plane (left) and the ratio between the relic density of $\psi$ and the total DM relic density in the $M_{\psi}$--$\Delta M$ plane (right), where $\Delta M = M_\psi - M_{W_D}$. All the points satisfy the relic constraint \eqref{eqn:DM-RD-3sigma} and pass the \texttt{HiggsBounds} and \texttt{HiggsSignal} checks.} 
    \label{fig:scatter-plot-1}
\end{figure}

In the LP of Fig.~\ref{fig:scatter-plot-1}, the spin-independent cross-section is shown in the $M_{h_2}$--$\sin\theta$ plane. One may see that, after taking into account all the relevant bounds, an upper bound is found for the Higgs mixing angle, $\sin\theta \lesssim 0.27$. We also see that the spin-independent cross-section has a weak dependence on the dark Higgs mass.
On the other hand, the spin-independent cross-section strongly depends on the Higgs mixing angle $\sin\theta$ which can be clearly seen from Eq.~\eqref{eqn:DD-expression}. Moreover, near to the SM Higgs resonance, there is a mutual cancellation between the SM Higgs and dark Higgs channel 
\cite{Baek:2011aa,Baek:2012se} which is clearly seen by the green points even for the higher values of the mixing angle $\sin\theta$.
In the RP of Fig.~\ref{fig:scatter-plot-1}, the ratio between the $\psi$ relic density and the total DM relic density is shown in the $M_\psi$--$\Delta M$ plane, where $\Delta M = M_\psi - M_{W_D}$. We find that if the mass gap between the initial- and final-state particles for the process $\psi \psi \rightarrow W_{D}W_{D}$ is small, then $\psi$ tends to contribute more to the DM relic density due to the phase space suppression. On the contrary, if the mass gap is large, then there will be less phase space suppression so the thermal average of cross-section times velocity will become large; this reduces $\psi$ relic density which is represented by the green points. The empty space in the top-left corner is due to the lower mass range of $W_D$ mass. 

\begin{figure}[t!]
    \centering
    \includegraphics[angle=0,height=7.5cm,width=7.5cm]{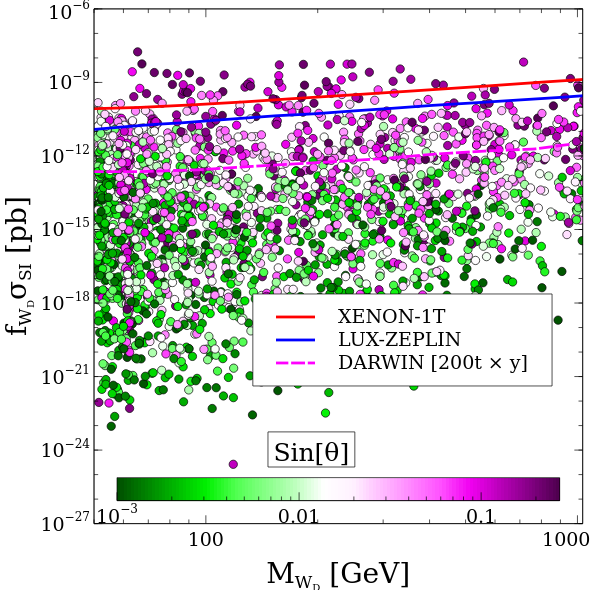}
    \includegraphics[angle=0,height=7.5cm,width=7.5cm]{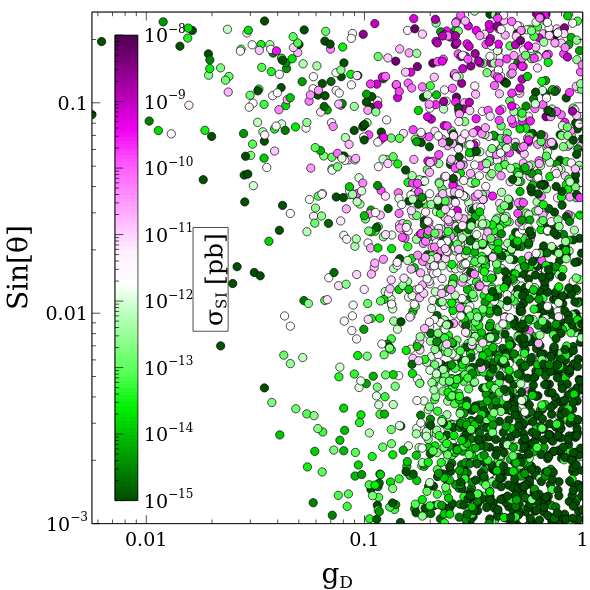}
    \caption{Scatter plots in the $M_{W_{D}}$--$\sigma_{\rm SI}$ plane (left) and $g_{D}$--$\sin\theta$ plane (right). All the points satisfy the relic constraint \eqref{eqn:DM-RD-3sigma} and pass the \texttt{HiggsBounds} and \texttt{HiggsSignal} checks. In the left panel, the colour bar represents the Higgs mixing angle, $\sin\theta$, 
    whereas in the right panel, it is the spin-independent direct detection cross-section, $\sigma_{\rm SI}$.} 
    \label{fig:scatter-plot-2}
\end{figure}

The LP and RP of Fig.~\ref{fig:scatter-plot-2} present scatter plots in the $M_{W_{D}}$--$\sigma_{\rm SI}$ and $g_{D}$--$\sin\theta$ planes, respectively. In the LP, the colour bar represents the value of $\sin\theta$. As can be seen from Eq.~\eqref{eqn:DD-expression}, $\sigma_{\rm SI}$ is proportional to the mixing angle $\sin^2 2\theta$, which is clearly visible from the colour variation in the figure. The recent LUX-ZEPLIN results \cite{LZ:2022lsv} already ruled out the $\sin\theta > 0.15$ region. 
Moreover, a large portion of the parameter space will be explored in the near future by the DARWIN experiment with its 200 tones $\times$ year exposure~\cite{DARWIN:2016hyl}, as depicted by the magenta dashed line in the LP of Fig.~\ref{fig:scatter-plot-2}.
In the RP, the colour bar depicts the spin-independent cross-section $\sigma_{\rm SI}$. We see that, once $g_{D}\gtrsim 0.08$ is considered, the $\sin\theta$ dependence becomes weaker. This happens because the dominant process in our setup is $W_{D}W_{D} \rightarrow h_{2}h_{2}$, when this process is kinematically allowed.
Then, the rate is proportional to $g^2_{D} (1-\sin^{2}\theta)$, which implies weaker dependence on $\sin\theta$.
Moreover, as we can see from Eq.~\eqref{eqn:DD-expression}, $\sigma_{\rm SI}$ depends on both $g_{D}$ and $\sin\theta$, and the transition of colour from green to magenta is observed if we increase either $g_{D}$ or $\sin\theta$. In principle, we also have DM annihilation like $W_{D}W_{D} \rightarrow {\rm SM}\,\, {\rm SM}$ mediated by $h_{1,2}$. However, if we lie outside the resonance region, we always overproduce DM. Having a parameter set in the exact resonance region, $M_{W_D} \sim M_{h_2}/2$, is less probable than having the $M_{W_D} > M_{h_2}$ case during the random scanning of the parameters~\eqref{eqn:random-scanning}. 
Therefore, the DM phenomenology for $W_{D}$ is mainly governed by the process $W_{D}W_{D} \rightarrow h_{2}h_{2}$. 

\begin{figure}[t!]
    \centering
    \includegraphics[angle=0,height=7.5cm,width=7.5cm]{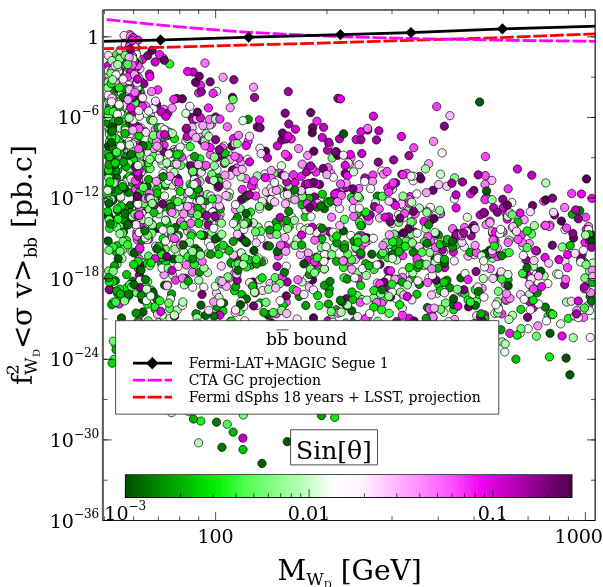}
    \includegraphics[angle=0,height=7.5cm,width=7.5cm]{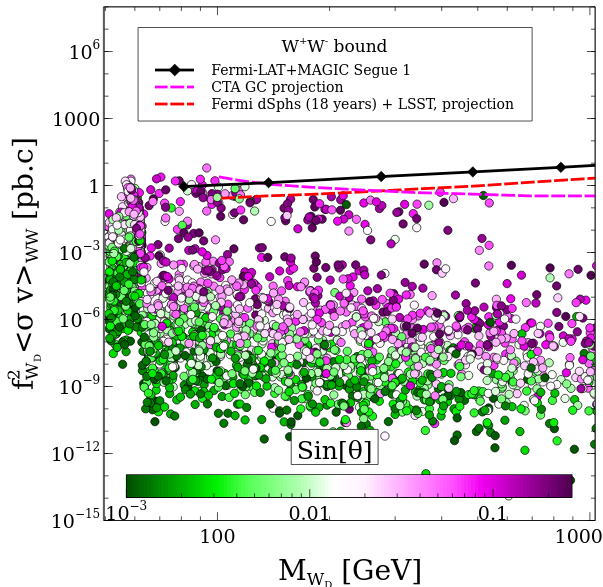}
    \caption{Scatter plots in the $M_{W_{D}}$--$\langle \sigma v\rangle_{b\bar{b}}$ (left) and $M_{W_{D}}$--$\langle \sigma v\rangle_{WW}$ (right) planes. The colour bars in both panels represent the Higgs mixing angle, $\sin\theta$. Here, the thermal averages of the cross-section times velocity, $\langle \sigma v\rangle$, are properly rescaled by $f_{W_D}^2 = (\Omega_{W_D}/\Omega_{\rm DM})^2$. All the points are obtained after imposing the relic constraint \eqref{eqn:DM-RD-3sigma} on the total DM relic density.} 
    \label{fig:scatter-plot-3}
\end{figure}

As we discussed above, the DM candidate $W_{D}$ mainly annihilates to $h_{2}h_{2}$, satisfying the constraint on the DM relic density. We thus expect that the DM annihilation to SM particles will be suppressed; otherwise, they would dominate the relic density. The LP of Fig.~\ref{fig:scatter-plot-3} shows the DM annihilation to $b\bar{b}$. The $y$-axis is the rescaled thermal average of the cross-section times velocity with $f_{W_D} = \Omega_{W_D} / \Omega_{\rm DM}$, and the $x$-axis is the mass of $W_D$. We find that the combined bound from Fermi-LAT and MAGIC~\cite{MAGIC:2016xys} on $b\bar{b}$ channel shown by the black line is ruling out a small region of the parameter space and for most of the parameter space, the bound is well above our predictions.
The prediction for $b\bar{b}$ from the Galactic Centre (GC) by the Cherenkov Telescope Array (CTA) \cite{CTA:2020qlo}, shown by the magenta dashed line, has already been explored partly by the Fermi-LAT and MAGIC observations. Moreover, the red dashed line represents the future sensitivity reach after combining Large Synoptic Survey Telescope (LSST) discoveries and continued data collection by Fermi-LAT for 18 years \cite{LSSTDarkMatterGroup:2019mwo}.
Future indirect detection experiments might explore the parameter space shown in the figure. Moreover, we see that the colour variation in $\sin\theta$ exhibits a linear correlation between $\langle \sigma v\rangle_{bb}$ and $M_{W_D}$, but there are also a few variations in the colour which happen due to the values of $g_D$ that can also alter the DM annihilation to the SM sector. The RP of Fig.~\ref{fig:scatter-plot-3} presents the DM annihilation to $W^+ W^-$. One may observe that the low-mass region of DM has already been explored by the Fermi-LAT and MAGIC $W^+W^-$ mode. We expect a little more parameter space to be explored in the future by the CTA \cite{CTA:2020qlo} and a combined analysis of LSST and Fermi-LAT \cite{LSSTDarkMatterGroup:2019mwo} as represented by the magenta dashed line and the red dashed line, respectively.

\begin{figure}[t!]
    \centering
    \includegraphics[angle=0,height=7.5cm,width=8.5cm]{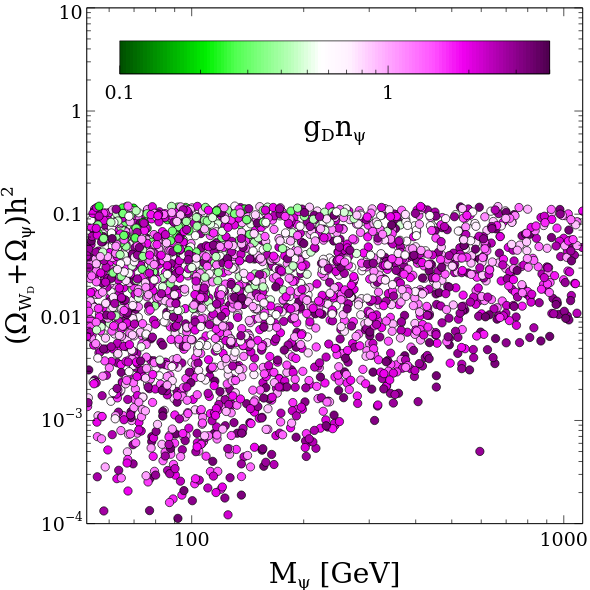}
    \caption{Total DM relic density, $\Omega_{\rm DM}h^2(=\Omega_{W_D}h^2 + \Omega_\psi h^2)$, in terms of the dark fermion mass $M_\psi$. The colour bar represents different values of the product of dark charge and gauge coupling $g_D n_\psi$.} 
    \label{fig:scatter-plot-4}
\end{figure}
Finally, Fig.~\ref{fig:scatter-plot-4} presents the total DM relic density, $\Omega_{\rm DM}h^2(=\Omega_{W_D}h^2 + \Omega_\psi h^2)$, in terms of the dark fermion mass $M_\psi$. The colour variation indicates different values of $g_D n_\psi$. We see that no points are allowed in the lower right corner due to the perturbative bound, $g_D n_\psi < \sqrt{4\pi}$. We note that small $g_D n_\psi$ values, represented by green points, give rise to the dominant contribution of the $\psi$ DM, whereas magenta points represent less contribution from the $\psi$ DM because of the large $g_D n_\psi$ values.
In general, $\psi$-DM relic density depends on the charge, gauge coupling, and its mass as $\Omega_\psi h^{2} \propto M_\psi^3/(g_D n_\psi)^4$. This behaviour is observed in Fig.~\ref{fig:scatter-plot-4}; for a fixed value of $M_\psi$, green points start to appear as we move to the larger relic density region. 
The presence of magenta points at $\Omega_{\rm DM}h^2 = 0.1226$ is due to the fact that, when the dark gauge boson DM $W_{D}$ contributes significantly, the $\psi$-DM contribution has to be small due to the upper bound on total DM relic density, {\it i.e.}, $\Omega_{\rm DM}h^2 = 0.1226$.

%%%%%%%%%%%%%%%%%%%%%%%%%%%%%%%%%%%%%%%%%%
\section{Inflation}
\label{sec:Inf}
%%%%%%%%%%%%%%%%%%%%%%%%%%%%%%%%%%%%%%%%%%

Having discussed in detail the DM phenomenology of the model, let us now move on to the possible realisation of cosmic inflation in the same model.
The action relevant for inflation is, in unitary gauge, given by
\begin{align}
    S = \int d^4x \, \sqrt{-g_{\rm J}} \, \left[
    \frac{M_{\rm P}^2}{2}\left(
    1 + \xi_H \frac{h^2}{M_{\rm P}^2} 
    + \xi_D \frac{\phi^2}{M_{\rm P}^2}
    \right)R_{\rm J}
    -\frac{1}{2}g_{\rm J}^{\mu\nu}\partial_\mu h \partial_\nu h
    -\frac{1}{2}g_{\rm J}^{\mu\nu}\partial_\mu \phi \partial_\nu \phi
    -V(\phi,h)
    \right]
    \,,\label{eqn:inflation_action_jordan}
\end{align}
where we have put the subscript J to denote that we are in the Jordan frame. The presence of the nonminimal couplings of the Higgs fields to the Ricci scalar, namely $\xi_H h^2 R$ and $\xi_D \phi^2 R$, may be seen natural as they have the mass-dimension of four. Moreover, even if we set the nonminimal couplings to zero at some energy scale, they will generically be generated through radiative corrections. In this work, we focus on positive nonminimal coupling parameters.
The scalar potential at tree level can be taken as
\begin{align}
    V(\phi,h) = \frac{1}{4}\lambda_H h^4 + \frac{1}{4}\lambda_D \phi^4 + \frac{1}{4}\lambda_{HD}\phi^2h^2\,,
    \label{eqn:inflaton_pot_jordan}
\end{align}
where we have omitted the quadratic mass terms which are negligible during inflation. We note that the model reduces to the standard Higgs inflation model \cite{Bezrukov:2007ep} in the $\phi\rightarrow 0$ limit.
Inflation with the Higgs-portal coupling is thoroughly studied in, {\it e.g.}, Ref.~\cite{Lebedev:2011aq} which we closely follow. We first discuss the aspect of inflation at the classical level, setting the notations. We then perform the quantum analysis, taking into account the suitable RG running of the coupling parameters.

%%%%%%%%%%%%%%%%%%%%%%%%%%%%%%%%%%%%%%%%%%
\subsection{Classical analysis}
\label{subsec:InfClassical}
%%%%%%%%%%%%%%%%%%%%%%%%%%%%%%%%%%%%%%%%%%
One may bring the Jordan-frame action \eqref{eqn:inflation_action_jordan} to the Einstein frame, denoted by the subscript E, via Weyl rescaling,
\begin{align}
    g_{{\rm J} \mu\nu} \rightarrow 
    g_{{\rm E} \mu\nu} = \Omega^2 g_{{\rm J} \mu\nu}
    \,,
\end{align}
with the conformal factor
\begin{align}
    \Omega^2 = 1 + \xi_H \frac{h^2}{M_{\rm P}^2}
    + \xi_D \frac{\phi^2}{M_{\rm P}^2}
    \,.
\end{align}
The Einstein-frame action is then obtained as
\begin{align}
    S = \int d^4x \, \sqrt{-g_{\rm E}} \, &\bigg[
    \frac{M_{\rm P}^2}{2}R_{\rm E}
    -\frac{3}{4}M_{\rm P}^2g_{\rm E}^{\mu\nu}\partial_\mu\ln\Omega^2\partial_\nu\ln\Omega^2
    \nonumber\\
    &\qquad\qquad
    -\frac{1}{2\Omega^2}g_{\rm E}^{\mu\nu}\partial_\mu h \partial_\nu h
    -\frac{1}{2\Omega^2}g_{\rm E}^{\mu\nu}\partial_\mu \phi \partial_\nu \phi
    -\frac{V}{\Omega^4}
    \bigg]\,.
\end{align}
Defining
\begin{align}
    \varphi \equiv \sqrt{\frac{3}{2}}M_{\rm P}\ln\Omega^2
    \,,\quad 
    \chi \equiv \frac{\phi}{h}
    \,,
\end{align}
we obtain
\begin{align}
    S = \int d^4x \, \sqrt{-g_{\rm E}} \, \left[
    \frac{M_{\rm P}^2}{2}R_{\rm E}
    -\frac{1}{2} \mathcal{K}_\varphi g_{\rm E}^{\mu\nu}\partial_\mu \varphi \partial_\nu \varphi 
    -\frac{1}{2} \mathcal{K}_\chi g_{\rm E}^{\mu\nu}\partial_\mu \chi \partial_\nu \chi 
    -\mathcal{K}_{\varphi\chi} g_{\rm E}^{\mu\nu}\partial_\mu \varphi \partial_\nu \chi 
    -U
    \right]\,,
\end{align}
where
\begin{align}
    \mathcal{K}_\varphi &=
    \frac{e^{\sqrt{\frac{2}{3}}\frac{\varphi}{M_{\rm P}}}(1+6\xi_H+(1+6\xi_D)\chi^2)-6(\xi_H+\xi_D\chi^2)}{6(\xi_H+\xi_D\chi^2)(e^{\sqrt{\frac{2}{3}}\frac{\varphi}{M_{\rm P}}}-1)}
    \,,\\
    \mathcal{K}_\chi &=
    \frac{M_{\rm P}^2(\xi_H^2+\xi_D^2\chi^2)}{(\xi_H+\xi_D\chi^2)^3}\left(1-e^{-\sqrt{\frac{2}{3}}\frac{\varphi}{M_{\rm P}}}\right)
    \,,\\
    \mathcal{K}_{\varphi\chi} &=
    \frac{M_{\rm P}(\xi_H-\xi_D)\chi}{\sqrt{6}(\xi_H+\xi_D\chi^2)^2}
    \,,
\end{align}
and $U$ is the Einstein-frame potential given by
\begin{align}
    U = \frac{V}{\Omega^4} = \frac{\lambda_H + \lambda_{HD}\chi^2 + \lambda_D\chi^4}{4(\xi_H+\xi_D\chi^2)^2}
    \left(
    1 - e^{-\sqrt{\frac{2}{3}}\frac{\varphi}{M_{\rm P}}}
    \right)^2
    M_{\rm P}^4
    \,.
\end{align}
We are interested in the large-field inflation where $\Omega^2 \gg 1$. In this limit,
\begin{align}
    \mathcal{K}_\varphi \approx
    1
    \,,\quad
    \mathcal{K}_\chi \approx 
    \frac{M_{\rm P}^2(\xi_H^2+\xi_D^2\chi^2)}{(\xi_H+\xi_D\chi^2)^3}
    \,,\quad
    \mathcal{K}_{\varphi\chi} \approx
    \frac{M_{\rm P}(\xi_H-\xi_D)\chi}{\sqrt{6}(\xi_H+\xi_D\chi^2)^2}
    \,.
\end{align}
Canonically normalising the $\chi$ field via
\begin{align}
    \left(\frac{d\chi_c}{d\chi}\right)^2 =
    \frac{M_{\rm P}^2(\xi_H^2+\xi_D^2\chi^2)}{(\xi_H+\xi_D\chi^2)^3}
    \,,
\end{align}
we obtain the effective action that is relevant for our consideration of inflation as follows:
\begin{align}
    S = \int d^4x \, \sqrt{-g_{\rm E}} \, &\bigg[
    \frac{M_{\rm P}^2}{2}R_{\rm E}
    -\frac{1}{2} g_{\rm E}^{\mu\nu}\partial_\mu \varphi \partial_\nu \varphi 
    -\frac{1}{2} g_{\rm E}^{\mu\nu}\partial_\mu \chi_c \partial_\nu \chi_c 
    \nonumber\\
    &\qquad
    -g_{\rm E}^{\mu\nu}\partial_\mu \varphi \partial_\nu \chi_c
    \frac{(\xi_H-\xi_D)\chi}{\sqrt{6}\sqrt{\xi_H^2+\xi_D^2\chi^2}\sqrt{\xi_H+\xi_D\chi^2}}
    -U
    \bigg]\,,\label{eqn:effective-action-inflation}
\end{align}
We note that for a finite, non-zero $\chi_c$, the kinetic mixing term vanishes when $\xi_H=\xi_D$. When $\xi_H \neq \xi_D$, the kinetic term gets suppressed for a large nonminimal coupling.
We are primarily interested in the case where at least one of the nonminimal couplings is large enough for us to safely ignore the kinetic mixing term.
Inflation may take along the SM Higgs direction, the dark Higgs direction, or the mixture of the SM and dark Higgs directions, which we call, respectively, SM Higgs inflation, dark Higgs inflation, and mixed inflation. We now look at each case in detail.

{\it (i) SM Higgs inflation scenario}: 
Let us first focus on the SM Higgs inflation scenario. The SM Higgs inflation corresponds to the $\chi=0$ case. We first note that
\begin{align}
    \frac{\partial U}{\partial \chi_c}\bigg\vert_{\chi=0} &= 0
    \,,\\
    \frac{\partial^2 U}{\partial \chi_c^2}\bigg\vert_{\chi=0} &=
    \frac{M^2(\lambda_{HD}\xi_H - 2\lambda_H\xi_D)}{2\xi_H^2}\left(
    1 - e^{-\sqrt{\frac{2}{3}}\frac{\varphi}{M_{\rm P}}}
    \right)^2\,.
\end{align}
Thus, $\chi=0$ becomes the minimum of the potential when
\begin{align}
    \lambda_{HD}\xi_H - 2\lambda_H\xi_D > 0\,,
\end{align}
or, equivalently, for $\xi_H>0$,
\begin{align}\label{eqn:SMHIconditiondv}
    \lambda_{HD} - 2\lambda_H\frac{\xi_D}{\xi_H} > 0
    \,.
\end{align}
Once the condition \eqref{eqn:SMHIconditiondv} is satisfied, we can work with the action,
\begin{align}
    S = \int d^4x \, \sqrt{-g_{\rm E}} \, &\bigg[
    \frac{M_{\rm P}^2}{2}R_{\rm E}
    -\frac{1}{2} g_{\rm E}^{\mu\nu}\partial_\mu \varphi \partial_\nu \varphi 
    -\frac{M_{\rm P}^4 \lambda_H}{4\xi_H^2}\left(
    1 - e^{-\sqrt{\frac{2}{3}}\frac{\varphi}{M_{\rm P}}}
    \right)^2
    \bigg]\,,
\end{align}
which coincides with the standard nonminimally-coupled single-field model, to find inflationary observables such as the spectral index $n_s$ and the tensor-to-scalar ratio $r$. In terms of the slow-roll parameters, defined as
\begin{align}
    \epsilon = \frac{M_{\rm P}^2}{2}\left(
    \frac{U'}{U}
    \right)^2
    \,,\quad
    \eta = M_{\rm P}^2 \frac{U''}{U}
    \,,\quad 
    \kappa^2 = M_{\rm P}^4\frac{U'U'''}{U^2}
    \,,
\end{align}
where the prime denotes the $\varphi$-field derivative, the spectral index and the tensor-to-scalar ratio are given, up to the second order in the slow-roll parameters, by \cite{Stewart:1993bc,Liddle:1994dx,Leach:2002ar}
\begin{align}
    n_s &=
    1 - 6\epsilon + 2\eta 
    - \frac{2}{3}(5+36c)\epsilon^2
    + 2(8c-1)\epsilon\eta + \frac{2}{3}\eta^2
    +\left(\frac{2}{3}-2c\right)\kappa^2
    \,,\label{eqn:SSI}\\
    r &=
    16\epsilon\left[
    1 + \left(
    4c-\frac{4}{3}
    \right)\epsilon + \left(
    \frac{2}{3} - 2c
    \right)\eta
    \right]
    \,,\label{eqn:TTSR}
\end{align}
while the amplitude of the curvature power spectrum is given by
\begin{align}\label{eqn:PSamp}
    A_s &= \frac{U}{24\pi^2M_{\rm P}^4\epsilon}\,,
\end{align}
where $c=\gamma + \ln 2 - 2$ with $\gamma \approx 0.5772$, and the quantities are understood to be evaluated at the horizon exit. The number of $e$-folds is given by
\begin{align}\label{eqn:efolds}
    N = -\frac{1}{M_{\rm P}^2}\int_{\varphi_*}^{\varphi_e} \frac{U}{U'} d\varphi \,,
\end{align}
where the subscript $e$ ($*$) denotes the end of inflation (the horizon exit).
For typical scenarios of reheating, we may take $N=60$ at the horizon exit. Using $\epsilon \simeq 1$ for the end of inflation, we can then get
\begin{align}\label{eqn:phiNSMHiggs}
    \varphi(N) \simeq \sqrt{\frac{3}{2}}M_{\rm P}\ln\left(\frac{4}{3}N\right)\,,
\end{align}
or, in terms of the original $h$ field,
\begin{align}
    h(N) \simeq \sqrt{\frac{4N}{3\xi_H}}M_{\rm P}\,.
\end{align}
Substituting Eq. \eqref{eqn:phiNSMHiggs} it into Eqs. \eqref{eqn:SSI} and \eqref{eqn:TTSR} gives
\begin{align}
    n_s &\approx
    1-\frac{2}{N}-\frac{19/6+2c}{N^2}
    -\frac{12c-3/2}{N^3}-\frac{15/4-27c}{2N^4}
    \,,\\
    r &\approx
    \frac{12}{N^2}-\frac{8(1-3c)}{N^3}-\frac{12(1-3c)}{N^4}\,.
\end{align}
For $N=60$, we find $n_s \approx 0.966$ and $r \approx 0.003$ which are in good agreement with the latest observational bounds \cite{Planck:2018jri,BICEP:2021xfz}.

{\it (ii) Dark Higgs inflation scenario}:
Let us now discuss the dark Higgs inflation scenario. The dark Higgs inflation corresponds to the $\chi=\infty$ case.
Noting that
\begin{align}
    \frac{\partial U}{\partial \chi_c}\bigg\vert_{\chi=\infty} &= 0
    \,,\\
    \frac{\partial^2 U}{\partial \chi_c^2}\bigg\vert_{\chi=\infty} &=
    \frac{M^2(\lambda_{HD}\xi_D - 2\lambda_D\xi_H)}{2\xi_D^2}\left(
    1 - e^{-\sqrt{\frac{2}{3}}\frac{\varphi}{M_{\rm P}}}
    \right)^2\,,
\end{align}
we see that, similar to the SM Higgs inflation case, $\chi=\infty$ is always an extremum. The $\chi=\infty$ direction becomes the minimum of the potential when
\begin{align}
    \lambda_{HD}\xi_D - 2\lambda_D\xi_H > 0\,,
\end{align}
or, equivalently, for $\xi_H>0$,
\begin{align}\label{eqn:DHIconditiondv}
    \lambda_{HD}\frac{\xi_D}{\xi_H} - 2\lambda_D > 0\,.
\end{align}
Once the condition \eqref{eqn:DHIconditiondv} is satisfied, we can work with the action,
\begin{align}
    S = \int d^4x \, \sqrt{-g_{\rm E}} \, &\bigg[
    \frac{M_{\rm P}^2}{2}R_{\rm E}
    -\frac{1}{2} g_{\rm E}^{\mu\nu}\partial_\mu \varphi \partial_\nu \varphi 
    -\frac{M_{\rm P}^4 \lambda_D}{4\xi_D^2}\left(
    1 - e^{-\sqrt{\frac{2}{3}}\frac{\varphi}{M_{\rm P}}}
    \right)^2
    \bigg]\,,
\end{align}
which becomes the same as the action for the SM Higgs inflation when we change the subscript $D$ to $H$. We can follow the same steps we performed above to find the spectral index and the tensor-to-scalar ratio. As the nonminimal coupling parameter and the quartic coupling parameter do not enter in the final expressions for $n_s$ and $r$, we conclude that we get the same prediction, namely $n_s \approx 0.966$ and $r \approx 0.003$.

{\it (iii) Mixed SM-dark Higgs inflation scenario}:
We finally consider the case where $\chi$ takes a finite, non-zero value, which we call $\chi_m$. In order for inflation to take places along the combined direction of $h$ and $\phi$ with $\chi=\chi_m$, we need the conditions,
\begin{align}
    \frac{\partial U}{\partial \chi_c}\bigg\vert_{\chi=\chi_m} = 0
    \,,\quad
    \frac{\partial^2 U}{\partial \chi_c^2}\bigg\vert_{\chi=\chi_m} > 0\,.
\end{align}
The first condition gives
\begin{align}
    (2\lambda_D \xi_H - \lambda_{HD} \xi_D)\chi_m^2
    -2\lambda_H \xi_D + \lambda_{HD} \xi_H = 0\,,
\end{align}
from which we find
\begin{align}\label{eqn:chimSq}
    \chi_m^2 = \frac{2\lambda_H \xi_D - \lambda_{HD} \xi_H}{2\lambda_D \xi_H - \lambda_{HD} \xi_D}\,.
\end{align}
Note that $\chi_m^2 > 0$ is required.
On the other hand, the second condition states
\begin{align}
    0 &<
    2\lambda_D\xi_H\chi_m^2(3\xi_H^3+2\xi_H\xi_D^2\chi_m^2-\xi_D^3\chi_m^4)
    -2\lambda_H\xi_D(\xi_H^3-2\xi_D\xi_H^2\chi_m^2-3\xi_D^3\chi_m^4)
    \nonumber\\
    &\quad
    +\lambda_{HD}(\xi_H^4-5\xi_D\xi_H^3\chi_m^2-5\xi_H\xi_D^3\chi_m^4+\xi_D^4\chi_m^6)
    \,,
\end{align}
or, upon using Eq.~\eqref{eqn:chimSq},
\begin{align}
    0 &< 
    \left(
    2\lambda_H\xi_D-\lambda_{HD}\xi_H
    \right)
    \left(
    \lambda_H\xi_D^2+\lambda_D\xi_H^2
    -\lambda_{HD}\xi_H\xi_D
    \right)
    \left[
    2\lambda_H\xi_D^3+2\lambda_D\xi_H^3
    -\lambda_{HD}\xi_H\xi_D(\xi_H+\xi_D)
    \right]
    \,.\label{eqn:mixedcondition}
\end{align}
For positive $\xi_H$ and $\xi_D$, the condition \eqref{eqn:mixedcondition} is equivalent to
\begin{align}\label{eqn:mixedconditiondv}
    2\lambda_H\frac{\xi_D}{\xi_H}-\lambda_{HD} > 0
    \,,\quad
    2\lambda_D-\lambda_{HD}\frac{\xi_D}{\xi_H} > 0
    \,.
\end{align}
Once the conditions $\chi_m^2>0$ and \eqref{eqn:mixedconditiondv} are satisfied, we can work with the action,
\begin{align}
    S = \int d^4x \, \sqrt{-g_{\rm E}} \, &\bigg[
    \frac{M_{\rm P}^2}{2}R_{\rm E}
    -\frac{1}{2} g_{\rm E}^{\mu\nu}\partial_\mu \varphi \partial_\nu \varphi 
    -\frac{M_{\rm P}^4 \lambda_m}{4\xi_m^2}\left(
    1 - e^{-\sqrt{\frac{2}{3}}\frac{\varphi}{M_{\rm P}}}
    \right)^2
    \bigg]\,,
\end{align}
where
\begin{align}
    \lambda_m &\equiv 
    4\lambda_H\lambda_D - \lambda_{HD}^2
    \,,\\
    \xi_m^2 &\equiv 
    4\lambda_D\xi_H^2 + 4\lambda_H\xi_D^2 - 4\lambda_{HD}\xi_H\xi_D 
    \,.
\end{align}
The action again becomes the same as the action for the SM Higgs inflation when we change the subscript $m$ to $H$. We get thus the same prediction, namely $n_s \approx 0.966$ and $r \approx 0.003$. 

\begin{figure}[ht!]
    \centering
    \includegraphics[scale=0.8]{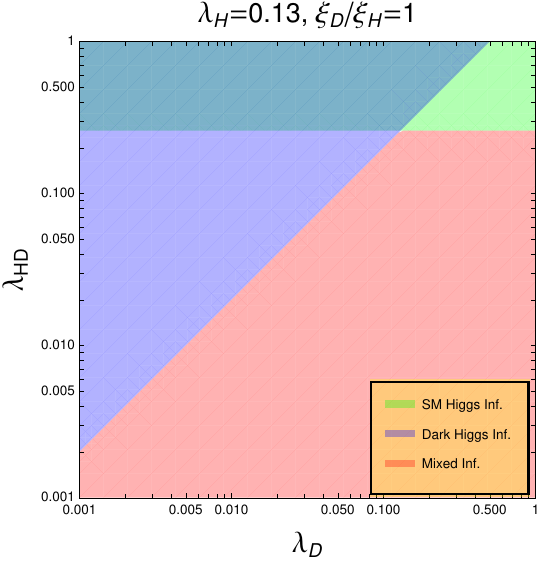}
    \includegraphics[scale=0.8]{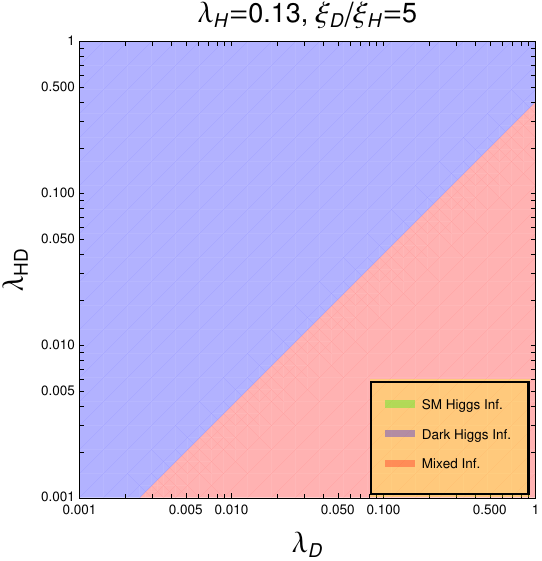}
    \\
    \includegraphics[scale=0.8]{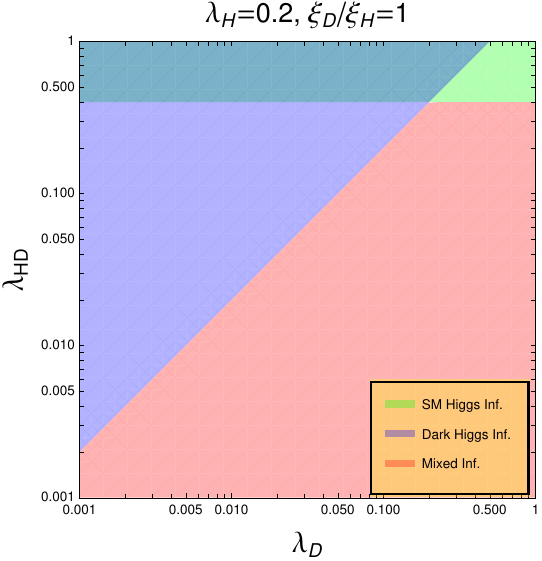}
    \includegraphics[scale=0.8]{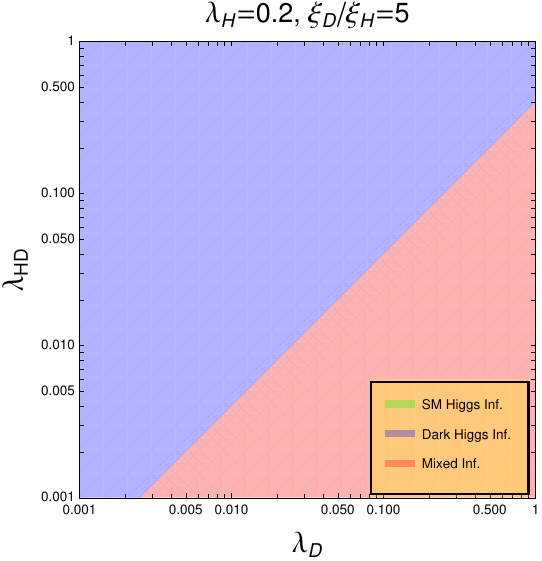}
    \caption{Allowed regions of classical SM Higgs inflation (green), dark Higgs inflation (blue), and mixed inflation (red), in the $\lambda_D$--$\lambda_{HD}$ plane. In the upper (lower) panel, $\lambda_H=0.13$ (0.2) is chosen, together with $\xi_D/\xi_H=1$ (left) and $\xi_D/\xi_H=5$ (right). As $\xi_D/\xi_H$ increases, the allowed region for the dark Higgs inflation expands. The allowed region for the SM Higgs inflation becomes smaller as $\lambda_H$ or $\xi_D/\xi_H$ takes a larger value. The allowed mixed inflation region shows a decreasing behaviour as $\xi_D/\xi_H$ increases on the plane we considered.}
    \label{fig:tree-level-case}
\end{figure}

In Fig.~\ref{fig:tree-level-case}, we present regions where inflation take place along the SM Higgs direction (green), the dark Higgs direction (blue), and the mixed direction (red) for $\{\lambda_H, \xi_D/\xi_H\}=\{0.13,1\}$, $\{0.13,5\}$, $\{0.2,1\}$, and $\{0.2,5\}$, in the $\lambda_D$--$\lambda_{HD}$ plane.
We observe that, as $\xi_D/\xi_H$ increases, the allowed region for the SM Higgs inflation shrinks, while the allowed region for the dark Higgs inflation expands. We also see that, as $\lambda_H$ increases, the allowed region for the SM Higgs inflation shrinks. While the allowed mixed inflation region shows a decreasing behaviour as $\xi_D/\xi_H$ increases on the plane we considered, it is not a universal tendency. From the mixed inflation condition \eqref{eqn:mixedconditiondv}, we see that the maximum value of $\lambda_{HD}$ is given by $2\lambda_H\xi_D/\xi_H$, which increases as $\lambda_H$ and $\xi_D/\xi_H$ become larger, while the lower bound on $\lambda_D$ for a given $\lambda_{HD}$, which is given by $(\xi_D/\xi_H)\lambda_{HD}/2$, becomes larger at the same time.

%%%%%%%%%%%%%%%%%%%%%%%%%%%%%%%%%%%%%%%%%%
\subsection{Quantum analysis}
\label{subsec:InfQuantum}
%%%%%%%%%%%%%%%%%%%%%%%%%%%%%%%%%%%%%%%%%%
In order to connect the high-energy scale of inflation to the low-energy scale of DM physics, it is vital to consider the RG running of the coupling parameters. For such a quantum analysis, we follow the procedures outlined in Ref.~\cite{Kim:2014kok} and consider the RG-improved effective action in the Jordan frame. In Sec.~\ref{subsec:InfClassical}, we have shown that the action relevant for inflation may effectively be given by a single-field action. We thus consider the following leading effective action:
\begin{align}
    \Gamma_{\rm eff} = \int d^4x \, \sqrt{-g_{\rm J}}\left[
    \frac{M_{\rm P}^2}{2} \Omega^2(t) R_{\rm J}
    -\frac{1}{2}g^{\mu\nu}_{\rm J}G^2(t)\partial_\mu\Phi(t)\partial_\nu\Phi(t)
    -V_{\rm eff}(t)
    \right]\,,
\end{align}
where $t=\ln(\mu/M_t)$, $\mu$ is the renormalisation scale, $M_t$ is the top-quark pole mass, and
\begin{align}
    \Omega^2(t) &= 1 + \xi_\Phi(t) G^2(t) \frac{\Phi^2(t)}{M_{\rm P}^2}
    \,,\\
    V_{\rm eff}(t) &= \frac{\lambda_\Phi(t)}{4}G^4(t)\Phi^4(t)
    \,,\\
    G(t) &= \exp\left(
    -\int^t dt' \, \frac{\gamma_\Phi}{1+\gamma_\Phi}
    \right)\,,
\end{align}
with $\Phi$ being the inflaton; for the SM (dark) Higgs inflation, $\Phi = h$ ($\Phi = \phi$), and for the mixed inflation case, $\Phi = \sqrt{1+\chi_m^2}h$ and
\begin{align}
    \xi_\Phi &= \frac{\xi_H + \xi_D \chi_m^2}{1+\chi_m^2}
    \,,\\
    \lambda_\Phi &= \frac{\lambda_H + \lambda_D\chi_m^4 + \lambda_{HD}\chi_m^2}{(1+\chi_m^2)^2}
    \,.
\end{align}
The RG equations as well as the anomalous dimensions are presented in Appendix~\ref{apdx:RGEs}.
In the Einstein frame, the effective action is given by
\begin{align}
    \Gamma_{\rm eff} = \int d^4x \, \sqrt{-g_{\rm E}}\left[
    \frac{M_{\rm P}^2}{2}R_{\rm E}
    -\frac{1}{2}g_{\rm E}^{\mu\nu}\partial_\mu\Psi(t)\partial_\nu\Psi(t)
    -U_{\rm eff}(t)
    \right]\,,
\end{align}
where $\Psi$ is the canonically-normalised field,
\begin{align}
    \left(
    \frac{\partial \Psi}{\partial \Phi}
    \right)^2 =
    \frac{G^2}{\Omega^2}
    +\frac{3M_{\rm P}^2}{2\Omega^4}\left(
    \frac{d\Omega^2}{d\Phi}
    \right)^2
    \,,
\end{align}
and the Einstein-frame effective potential is
\begin{align}\label{eqn:EFeffPot}
    U_{\rm eff}(t) =
    \frac{\lambda_\Phi(t)G^4(t)\Phi^4(t)}{4(1+\xi_\Phi(t)G^2(t)\Phi^2(t)/M_{\rm P}^2)^2}\,.
\end{align}

The conditions for the SM Higgs inflation, dark inflation, and the mixed inflation are the same as Eqs. \eqref{eqn:SMHIconditiondv}, \eqref{eqn:DHIconditiondv}, and \eqref{eqn:mixedconditiondv}, respectively. The only difference is that the conditions should be met at the inflation scale.
The scheme we use to find the inflation scale is as follows (see also Ref.~\cite{Kim:2014kok}).
The package \texttt{mr} \cite{Kniehl:2015nwa,Kniehl:2016enc} is utilised to read the $\overline{\rm MS}$-running, EW parameters at the top-quark pole mass $M_t$ with the latest PDG values \cite{ParticleDataGroup:2022pth}, $M_t=172.5$ GeV, $M_W=80.377$ GeV, $M_Z=91.1876$ GeV, $M_{h_1}=125.25$ GeV, $G_F=1.1664\times 10^{-5}$ GeV${}^{-2}$, $\alpha=1/127.951$, and $\alpha_s=0.118$, where $G_F$ is the Fermi coupling constant, $\alpha$ is the fine-structure constant at $M_Z$, and $\alpha_s$ is the strong coupling constant at $M_Z$; see also Ref.~\cite{Buttazzo:2013uya}.
For a given set of parameters at $\mu=M_t$, $\{\lambda_H, \lambda_{HD}, \lambda_D, n_\psi, g_D, \xi_D/\xi_H\}$, we run to the Planck scale using the RG equations presented in Appendix~\ref{apdx:RGEs}. 
One may notice from the beta function expressions in Appendix~\ref{apdx:RGEs} that the Higgs-portal coupling $\lambda_{HD}$ positively contributes to the running of the SM Higgs quartic coupling $\lambda_H$. Thus, the instability problem of the SM Higgs potential can be lifted, and the SM Higgs quartic coupling can stay positive up to the inflation scale as advertised in the introduction.
We find the scale for end of inflation through the condition $\epsilon = 1$ and the horizon-exit scale by imposing 60 $e$-folds.
We then examine the inflation conditions \eqref{eqn:SMHIconditiondv}, \eqref{eqn:DHIconditiondv}, and \eqref{eqn:mixedconditiondv} together with the perturbativity conditions as well as the instability conditions, $\lambda_\Phi > 0$. Note that one of the nonminimal coupling parameters is not a free parameter. Rather, it is given by the normalisation condition that the amplitude of the curvature power spectrum \eqref{eqn:PSamp} is $A_s \simeq 2.1 \times 10^{-9}$ at the pivot scale; for the SM Higgs inflation, for instance, $\xi_H$ shall be fixed in this manner, leaving only the ratio $\xi_D/\xi_H$ as a free input parameter.

\begin{figure}[t!]
    \centering
    \includegraphics[scale=0.8]{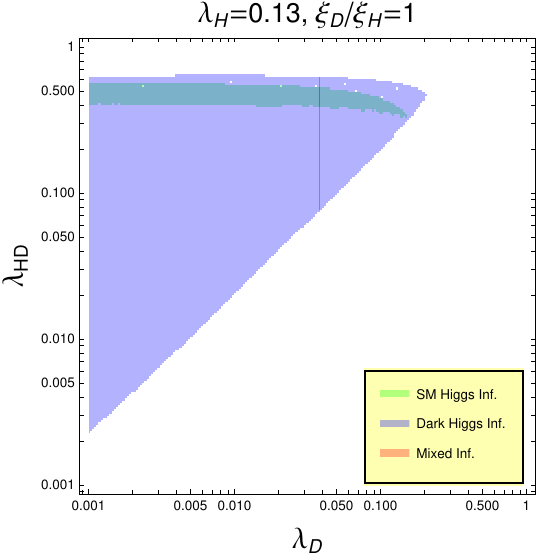}
    \includegraphics[scale=0.8]{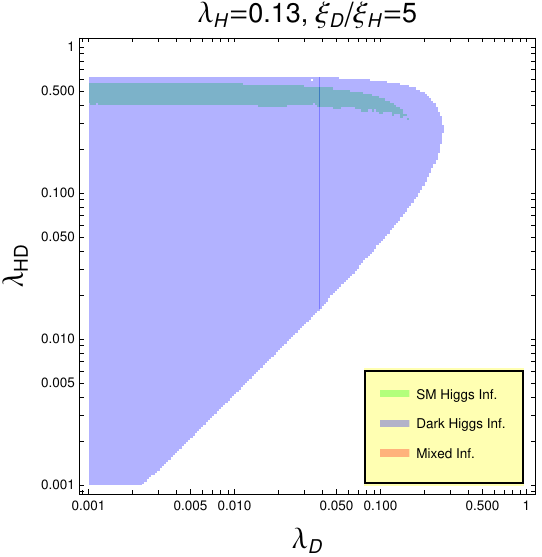}
    \\
    \includegraphics[scale=0.8]{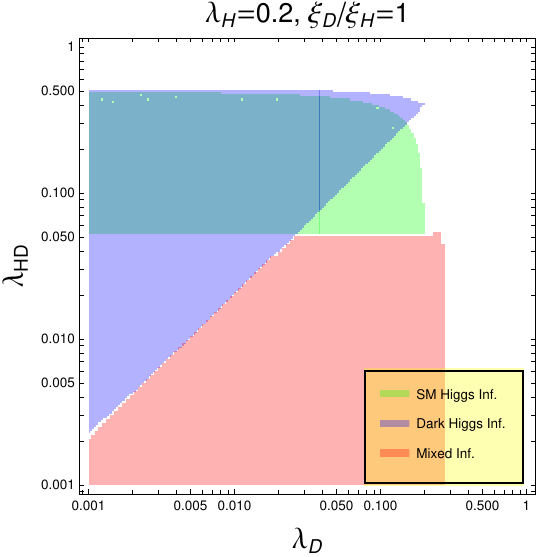}
    \includegraphics[scale=0.8]{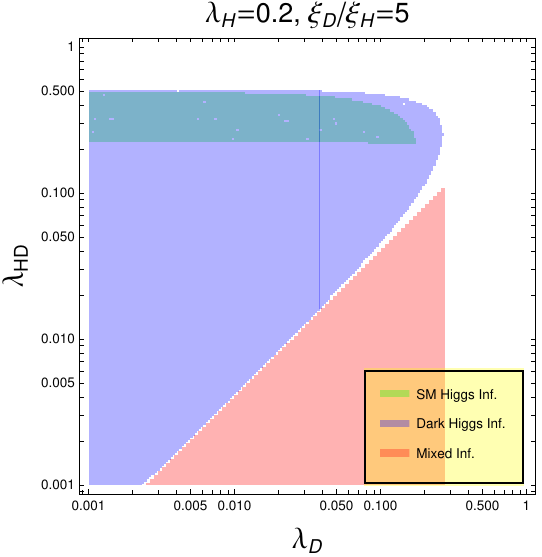}
    \caption{Allowed regions of SM Higgs inflation (green), dark Higgs inflation (blue), and mixed inflation (red), in the $\lambda_D$--$\lambda_{HD}$ plane at the inflation scale after taking into account the RG running. In the upper (lower) panel, $\lambda_H=0.13$ (0.2) is chosen, together with $\xi_D/\xi_H=1$ (left) and $\xi_D/\xi_H=5$ (right). For all the cases, we have chosen $n_\psi=1$ and $g_D=0.1$. See the main text for all the conditions imposed. The SM Higgs inflation becomes more viable as $\lambda_H$ increases. The allowed region of the dark Higgs inflation expands as $\xi_D/\xi_H$ takes a larger value. In the parameter space we considered here, no mixed inflation is possible with $\lambda_H=0.13$.}
    \label{fig:quantum-case}
\end{figure}

The results are shown in Fig.~\ref{fig:quantum-case}. The allowed regions of the SM Higgs inflation (green), dark Higgs inflation (blue), and mixed inflation (red) are presented in the $\lambda_D$--$\lambda_{HD}$ plane. Similar to the classical cases, we have considered $\{\lambda_H, \xi_D/\xi_H\}=\{0.13,1\}$, $\{0.13,5\}$, $\{0.2,1\}$, and $\{0.2,5\}$, while fixing $n_\psi=1$ and $g_D=0.1$.
We stress that the input parameters are chosen at the top-quark pole mass $M_t$. In particular, the choice of $\xi_D/\xi_H=1$, in which case the kinetic mixing term vanishes as we mentioned below Eq.~\eqref{eqn:effective-action-inflation}, is given at $M_t$ and does not hold at all scales due to the RG running; see Appendix~\ref{apdx:RGEs}.
As stated before, one of the nonminimal couplings is fixed to match $A_s \simeq 2.1\times 10^{-9}$ at the inflation scale. For instance, in the case of the scan presented in Fig.~\ref{fig:quantum-case}, the value of $\xi_H$ for the SM Higgs inflation falls in the region $\xi_H=\{975.81, 8603.5\}$ (top-left), $\{558.69, 7604.8\}$ (top-right), $\{9245.4, 14495\}$ (bottom-left), and $\{4838.1, 10299\}$ (bottom-right). It agrees with our understanding that smaller values of $\lambda_H$ at the inflation scale requires smaller values of $\xi_H$.\footnote{
It is also possible to make $\lambda_H$ take an extremely small value at the inflation scale by tuning the input parameters, thereby further reducing the value of $\xi_H$. We do not consider such a highly fine-tuned case in this work.
} While $\xi_H\gg 1$ causes the unitarity issue for the SM Higgs inflation, the inflationary analysis is not compromised as the unitarity violation scale is higher than the inflation scale \cite{Bezrukov:2010jz}; see also our discussion in the introduction. The viable SM Higgs inflation region becomes larger as we increase $\lambda_H$. It is mainly due to the fact that the instability of $\lambda_H<0$ disappears. The allowed dark Higgs inflation region expands as $\xi_D/\xi_H$ takes a larger value. This is similar to the classical case. On the parameter space we considered here, no mixed inflation is possible with $\lambda_H=0.13$. One may easily notice that the quantum analysis gives a very different result from the classical analysis shown in Fig.~\ref{fig:tree-level-case}, and thus, it is crucial to properly take into account the quantum effects when attempting to make connections with DM physics.

\begin{figure}[t!]
    \centering
    \includegraphics[scale=0.5]{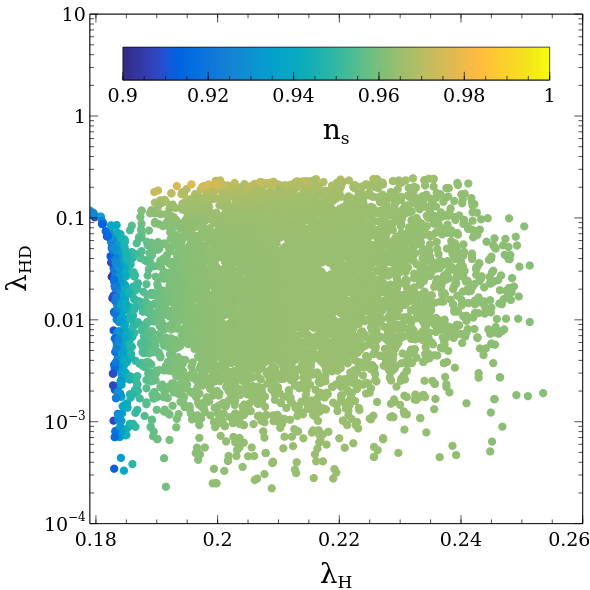}
    \includegraphics[scale=0.5]{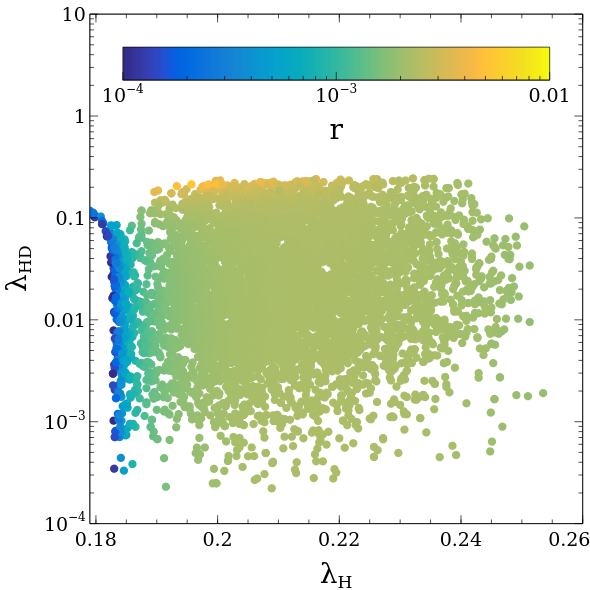}
    \caption{Scalar spectral index $n_s$ (left) and tensor-to-scalar ratio $r$ (right) in the $\lambda_H$--$\lambda_{HD}$ plane. While the spectral index may take a very small number of 0.9 or a large number of 1, in the wide range of the parameter space, $n_s \simeq 0.96$, which is preferred by the latest observational data. We observe that the tensor-to-scalar ratio always remains to be smaller than 0.01, being compatible with the latest observational bound.}
    \label{fig:scatter-plot1}
\end{figure}

With the Einstein-frame potential \eqref{eqn:EFeffPot}, we can compute inflationary observables such as the spectral index $n_s$ and the tensor-to-scalar ratio $r$ as sketched in Sec.~\ref{subsec:InfClassical}. 
Let us consider the SM Higgs inflation. In order to compute $n_s$ and $r$ for a wide range of parameter sets, we scan over 
\begin{align}\label{eqn:NM-scan}
10^{-3} \leq \frac{\xi_D}{\xi_H} \leq 10\,,
\end{align}
in addition to Eq.~\eqref{eqn:random-scanning}.
When performing the random scan, in addition to the SM Higgs inflation condition \eqref{eqn:SMHIconditiondv}, we have also demanded that the quartic couplings stay positive up to the Planck scale, {\it i.e.}, the stability condition, and always less or equal to $4\pi$, {\it i.e.}, the perturbativity condition.
The results for $n_s$ and $r$ are presented in Fig.~\ref{fig:scatter-plot1} in the $\lambda_H$--$\lambda_{HD}$ plane. From the LP, we see that $\lambda_H \sim 0.19$ is in tension with the latest observational bound, $0.958 \leq n_s \leq 0.975$ (95\% C.L.) \cite{Planck:2018jri,BICEP:2021xfz}, as the spectral index becomes too small. The region $\lambda_{HD} > 0.1$ is also in tension with the bound due to higher values of $n_s$. In the wide range of parameter space, however, we obtain $n_s \simeq 0.96$ which is preferred by the latest observations.
The tensor-to-scalar ratio is shown in the RP. We observe that the tensor-to-scalar ratio always remains to be smaller than 0.01, which is well within the latest observational bound of $r\leq 0.036$ (95\% C.L.) \cite{Planck:2018jri,BICEP:2021xfz}. 

\begin{figure}[t!]
    \centering
    \includegraphics[scale=0.6]{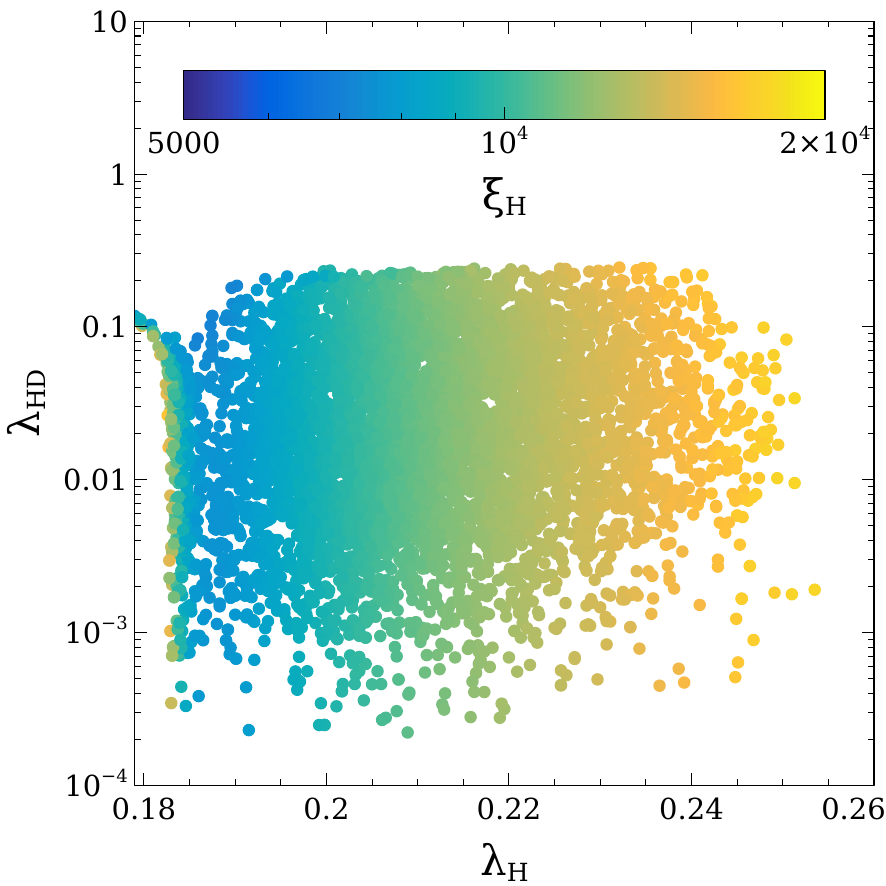}
    \caption{Nonminimal coupling $\xi_H$ in the $\lambda_H$--$\lambda_{HD}$ plane. Points correspond to the same scanned points presented in Fig.~\ref{fig:scatter-plot1}. For a given input parameter set, the nonminimal coupling $\xi_H$ is obtained by requiring that the amplitude of the curvature power spectrum matches $A_s \simeq 2.1 \times 10^{-9}$ at the inflation scale.}
    \label{fig:scatter-plot-xiH}
\end{figure}
For the same scanned points presented in Fig.~\ref{fig:scatter-plot1}, values of the nonminimal coupling $\xi_H$ are shown in Fig.~\ref{fig:scatter-plot-xiH}. We see that the nonminimal coupling $\xi_H$, which is obtained by requiring that the amplitude of the curvature power spectrum matches $A_s \simeq 2.1 \times 10^{-9}$ at the inflation scale, takes values of $\xi_H\sim\mathcal{O}(10^{4})$. 
As the nonminimal coupling is much larger than unity, $\xi_H \gg 1$, we anticipate the unitarity violation. Nevertheless, as the inflation scale is well below the unitarity violation scale, the computation of the inflationary observables such as the scalar spectral index $n_s$ and the tensor-to-scalar ratio $r$ is still credible.
From Fig.~\ref{fig:scatter-plot-xiH}, one can also observe that larger values of the SM Higgs quartic coupling $\lambda_H$ require larger values of the SM Higgs nonminimal coupling $\xi_H$. This behaviour agrees well with our finding in Fig.~\ref{fig:quantum-case} and with our understanding that $\xi_H \propto \sqrt{\lambda_H}$ discussed in the introduction.

%%%%%%%%%%%%%%%%%%%%%%%%%%%%%%%%%%%%%%%%%%
\section{Correlations between inflation and dark matter}
\label{sec:DMInf}
%%%%%%%%%%%%%%%%%%%%%%%%%%%%%%%%%%%%%%%%%%

In this section, we investigate the allowed parameter space, by performing a scan over the range \eqref{eqn:random-scanning} and \eqref{eqn:NM-scan}, after imposing the various bounds from DM as well as from inflation. In the case of DM, we mainly use the DM relic density bound, DM direct detection bound, and indirect detection bound. Moreover, we also use the bounds on the Higgs sector using \texttt{HiggsBound} and \texttt{HiggsSignal}.
For inflation, one may consider either the SM Higgs inflation, dark Higgs inflation, or the mixed inflation. While different choices of inflation would result in different allowed parameter spaces, all the three scenarios are qualitatively equivalent. From the viewpoint of the RG running, the SM Higgs inflation may be considered hard to realise as, in the pure SM, the SM Higgs quartic coupling becomes negative, {\it i.e.}, instability, before reaching the inflation scale.
Therefore, in this work, we focus on the SM Higgs inflation scenario, leaving detailed analyses for the other two scenarios as future work. We impose the bounds on the spectral index $n_s$ and the scalar-to-tensor ratio $r$, together with the inflation condition \eqref{eqn:SMHIconditiondv}.

\begin{figure}[t!]
    \centering
    \includegraphics[scale=0.5]{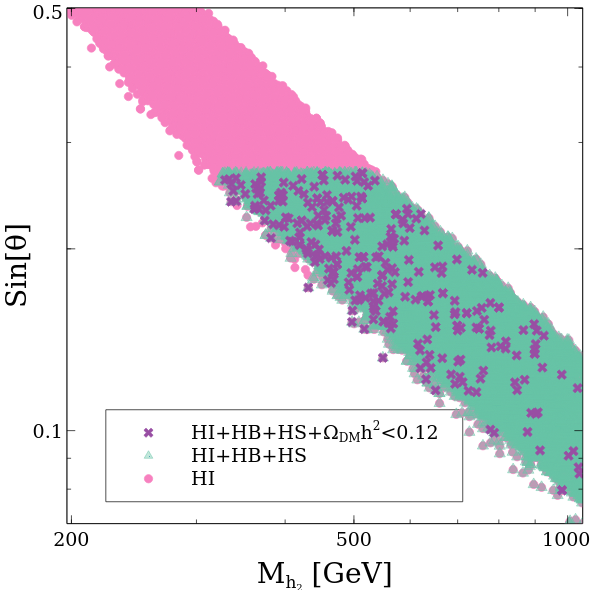}
    \includegraphics[scale=0.5]{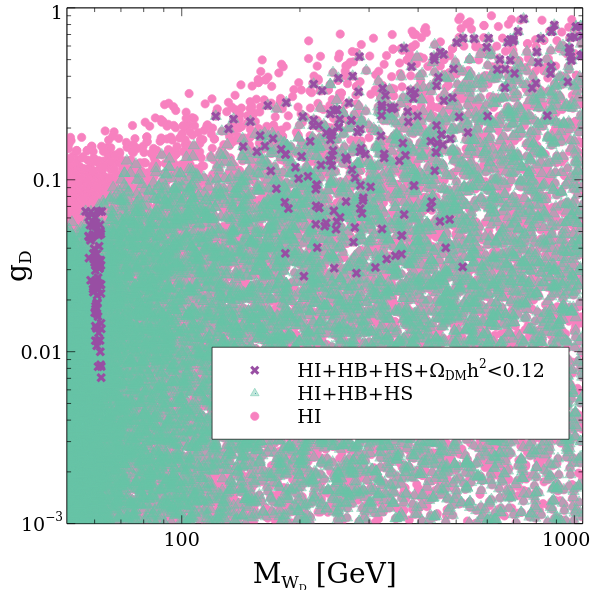}
    \caption{Allowed parameter space in the $M_{h_{2}}$--$\sin[\theta]$ (left) and $M_{W_{D}}$--$g_{D}$ (right) planes. All the points shown here satisfy $\Omega_{\rm DM}h^2 > 10^{-4}$.
    Together with the inflation-related bounds (referred to as HI), we obtain the magenta, circle points. The green, triangle points are obtained after imposing bounds on the strength of the couplings associated with the SM Higgs and dark Higgs using \texttt{HiggsSignal} (referred to as HS) and \texttt{HiggsBounds} (referred to as HB) together with the inflation-related bounds. Finally, the violet, cross points are obtained after further demanding the DM relic density to be less than the total density of the Universe, {\it i.e.}, $\Omega_{\rm DM}h^2 \leq 0.12$.}
    \label{fig:scatter-plot2}
\end{figure}
In the LP of Fig.~\ref{fig:scatter-plot2}, the allowed parameter space is shown in the $M_{h_{2}}$--$\sin\theta$ plane. All the points satisfy $\Omega_{\rm DM}h^2 > 10^{-4}$, the perturbativity condition, and the inflation-related bounds (referred to as HI in the plot) such as $0.958 \leq n_s \leq 0.975$, $r\leq 0.036$, the instability condition, and the SM Higgs inflation condition \eqref{eqn:SMHIconditiondv}. 
Imposing the bound associated with the Higgs sector using \texttt{HiggsSignal} (denoted by HS) and \texttt{HiggsBound} (denoted by HB) on top of HI leaves us the green, triangle points. The violet, cross points are obtained after further imposing the upper limit of the DM relic density $\Omega_{\rm DM}h^{2} \leq 0.12$. 
We see a nice correlation between $M_{h_2}$ and the mixing angle $\sin\theta$. This is because we need a relatively large SM Higgs quartic coupling in order to avoid it becoming negative at high-energy scales. The relation $\lambda_H \propto \sin^{2}\theta M_{h_2}^2$ is exactly the behaviour we observe in the plot. The bounds associated with the SM Higgs precision data and dark Higgs exclude the small $h_2$ mass, $M_{h_2} < 300$ GeV, and large Higgs mixing angle, $\sin\theta > 0.27$. This region is shown by the green, triangle points. 
Finally, imposing the upper bound on the DM relic density $\Omega_{\rm DM} h^{2} \leq 0.12$ excludes further points, leaving us the violet, cross points.

The RP of Fig.~\ref{fig:scatter-plot2} presents the allowed parameter space in the $M_{W_D}$--$g_{D}$ plane. One may notice that a large region is allowed if only the inflation-related bounds are imposed. If we screen the points by further using \texttt{HiggsBound} and \texttt{HiggsSignal}, we are left with the green, triangle points. We see that for $M_{W_D} < 200$ GeV and $g_{D} > 0.1$, a part of the region gets ruled out. This is because the region below $M_{W_D} < M_{h_1}/2$ is disfavoured mostly due to the Higgs invisible decay width, and beyond this kinematical limit, it is due to the SM Higgs signal strength. The violet, cross points are obtained after imposing the DM relic density limit as well. We see a line at $M_{W_D} \sim M_{h_1}/2$ which implies the SM Higgs resonance region; due to the presence of the Higgs resonance, those points are allowed from the DM relic density bound. A gap in $M_{h_1}/2 \lesssim M_{W_D} \lesssim M_{h_1}$ is because of the over-production of DM. Once the $W_{D}W_{D} \rightarrow h_{1}h_{1}$ channel opens, we start getting points again; as the mass of $W_D$ increases, we get more points due to the resonance associated with $h_2$. We also observe that low values of $g_D$ are disfavoured; this is due to the over-production of DM.

\begin{figure}[t!]
    \centering
    \includegraphics[scale=0.5]{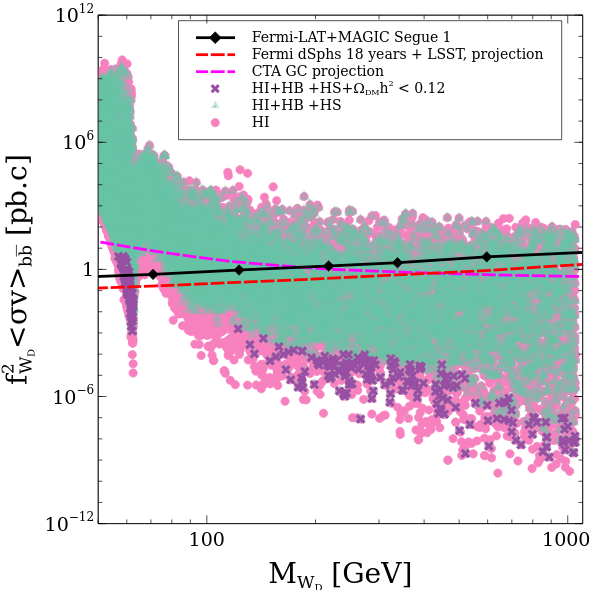}
    \includegraphics[scale=0.5]{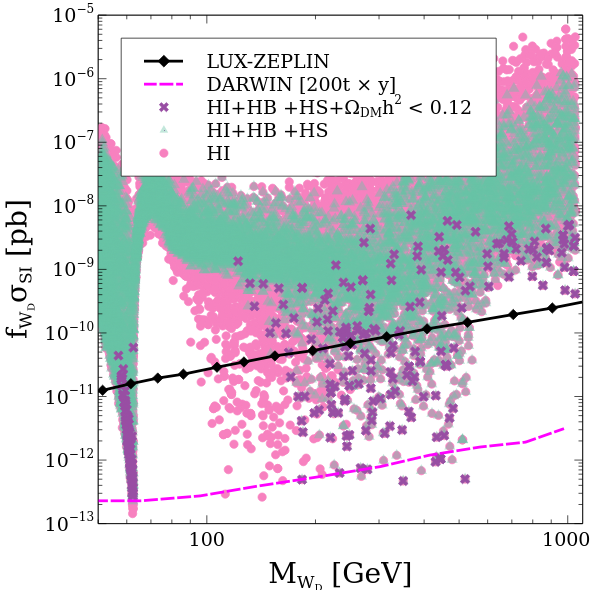}
    \caption{$f^2_{W_{D}} \langle\sigma v\rangle_{b\bar{b}}$ (left) and $f_{W_{D}} \sigma_{\rm SI}$ (right) in terms of $M_{W_{D}}$. The black lines indicate the bounds from Fermi-LAT (left) and LUX-ZEPLIN (right). The magenta dashed lines represents the future projection by the CTA (left) and by the DARWIN (right). Finally, the red dashed line on the LP indicates the future sensitivity after combining LSST discoveries and continued data collection by Fermi-LAT for 18 years.
    The same colour scheme is adopted as in Fig.~\ref{fig:scatter-plot2}. All the points satisfy $\Omega_{\rm DM}h^{2} > 10^{-4}$.}
    \label{fig:scatter-plot3}
\end{figure}

In the LP and RP of Fig.~\ref{fig:scatter-plot3}, we present scatter plots in the $M_{W_D}$--$f^2_{W_D} \langle\sigma v\rangle_{b\bar{b}}$ and $M_{W_D}$--$f_{W_D} \sigma_{\rm SI}$ planes, respectively.
In the LP, we see that a small part near the SM Higgs resonance region is in conflict with the Fermi-LAT indirect detection bound for the $b\bar{b}$ channel. The region $M_{W_D} > M_{h_1}$ is, on the other hand, well below the present Fermi-LAT bound.
We have also shown, with the magenta dashed line, the future projection by the CTA aiming to study the DM signal from the GC \cite{CTA:2020qlo}. On the other hand, the red dashed line represents the future projection after combining the LSST discoveries with continued data collection of Fermi-LAT for 18 years \cite{LSSTDarkMatterGroup:2019mwo}.
The RP shows a rescaled spin-independent cross-section together with the severe bound of LUX-ZEPLIN. We see that a small portion near the SM Higgs resonance and a large portion of larger $M_{W_D}$ ranges are already ruled out by the direct detection bound imposed by the LUX-ZEPLIN data. 
Moreover, the full parameter space will be explored by the DARWIN experiment in the future with its 200 tones $\times$ year exposure \cite{DARWIN:2016hyl}.

%%%%%%%%%%%%%%%%%%%%%%%%%%%%%%%%%%%%%%%%%%
\section{Conclusion}
\label{sec:conc}
%%%%%%%%%%%%%%%%%%%%%%%%%%%%%%%%%%%%%%%%%%

Considering a dark $U(1)_D$ extension of the Standard Model, we have investigated dark matter phenomenology. In addition to the Standard Model fields, the model includes three dark fields, namely a dark Higgs field, a dark fermion, and a dark vector boson that is associated with the dark $U(1)_D$. The dark fermion and the dark vector boson naturally become dark matter candidates, and thus, the model could feature a two-component dark matter scenario. We have performed a detailed numerical analysis with various constraints such as the dark matter relic density, collider bounds, as well as direct and indirect detection bounds to identify allowed regions in the parameter space. We have found that a large portion of the parameter space may accommodate the two-component dark matter scenario. In particular, the fermion dark matter $\psi$ is not accessible by direct detection experiments. Therefore, negative results from dark matter direct detection experiments do not necessarily mean that WIMP scenario is strongly disfavoured. Our model is a proof of existence for such a case.

We have also investigated the possibility of realising cosmic inflation in the same model. As the model contains two scalar fields, inflation may be realised as three different scenarios: the Standard Model Higgs inflation, the dark Higgs inflation, and the mixed case. We have first analysed at the classical level all these three scenarios with the inclusion of nonminimal coupling terms and identified the parameter space in which these scenarios could be realised. We have then performed the quantum analysis by utilising the renormalisation group running of the coupling parameters and the renormalisation group-improved effective action, focusing on the Standard Model Higgs inflation scenario. We have found that a small portion of the parameter space becomes incompatible with the latest observational bounds as the spectral index becomes either too small or too large. However, a wide range of the parameter space resulted in spectral index values that sit within the allowed bound. The tensor-to-scalar ratio turned out to be always smaller than the current upper limit.

Through the running of the coupling parameters, the high-energy scale physics of inflation could be connected to the low-energy scale physics of dark matter. We have performed a thorough scan with the imposition of both the dark matter-related constraints and the inflation-related constraints. We have found that while the model is capable of accommodating both dark matter and inflation, the model becomes tightly constrained, and only a small section of the parameter space survives. As more and more observational data become accessible, and with future experiments, we expect to test the remaining small section of the parameter space. We may be able to even rule out the model entirely as a unified framework for both dark matter and the Standard Model Higgs inflation.

While we have examined the allowed parameter space for the three inflation scenarios, we have paid extra attention to the Standard Model Higgs inflation case when connecting inflation to dark matter. However, it is certainly possible that inflation takes place along the dark Higgs field or the direction of the combination of the Standard Model Higgs and the dark Higgs fields. The other inflation scenarios may open up the allowed parameter regions. We plan to explore these possibilities in the future.

%%%%%%%%%%%%%%%%%%%%%%%%%%%%%%%%%%%%%%%%%%
\acknowledgments

This work used the Scientific Compute Cluster at GWDG, the joint data center of Max Planck Society for the Advancement of Science (MPG) and University of G\"{o}ttingen.
PK is supported in part by the KIAS Individual Grant No. PG021403, and by National Research Foundation of Korea (NRF) Research Grant NRF-2019R1A2C3005009. 
%%%%%%%%%%%%%%%%%%%%%%%%%%%%%%%%%%%%%%%%%%

\appendix

%%%%%%%%%%%%%%%%%%%%%%%%%%%%%%%%%%%%%%%%%%
\section{Cross-section expressions}
\label{apdx:XSecs}
%%%%%%%%%%%%%%%%%%%%%%%%%%%%%%%%%%%%%%%%%%
We summarise expressions for the cross-sections relevant for our study.

\begin{itemize}
\item {\bf $ W_{D}W_{D} \rightarrow W^{+}W^{-}:$}

The cross-section takes 
\begin{align}
    \sigma = \frac{1}{16 \pi s} \left( \frac{s-4M^2_{W}}{s-4M^2_{W_D}} \right)^{1/2}\, |M|^2_{WW}
    \,,
\end{align}
where $s$ is the Mandelstam variable, and the amplitude $|M|_{WW}$ is expressed as
\begin{align}
    |M|^{2}_{WW} &= \frac{4}{9}  \left| \frac{g_{h_{1}W_{D}W_{D}} g_{h_{1}WW }}{(s-M^2_{h_1}) 
    + i \Gamma_{h_1} M_{h_1}} + \frac{g_{h_{2}W_{D}W_{D}} g_{h_{2}WW }}{(s-M^2_{h_2}) 
    + i \Gamma_{h_2} M_{h_2}} \right|^{2} 
    \nonumber \\
    &\quad
    \times \left( 1 + \frac{(s-2M^2_{W_{D}})^2}{8 M^4_{W_{D}}}\right) \left( 1 + \frac{(s-2M^2_{W})^2}{8 M^4_{W}}\right)\,.
\end{align}
The vertices are given by
\begin{align}
    g_{h_{1(2)W_{D}W_{D}}} &= -2 g_{D} M_{W_{D}} \sin \theta (-\cos\theta)
    \,,\nonumber \\
    g_{h_{1(2)WW}} &= \frac{v}{2 s^2_w} \cos \theta\, (\sin\theta)
    \,,
\end{align}
where $s^2_w = 0.23$ is the Weinberg angle.

\item {\bf $ W_{D}W_{D} \rightarrow ZZ:$}

The cross-section takes
\begin{align}
    \sigma = \frac{1}{32 \pi s} \left( \frac{s-4M^2_{Z}}{s-4M^2_{W_D}} \right)^{1/2}\, |M|^2_{ZZ}\,.
\end{align}
The amplitude $|M|_{WW}$ is expressed as
\begin{align}
    |M|^{2}_{ZZ} &= \frac{4}{9}  \left| \frac{g_{h_{1}W_{D}W_{D}} g_{h_{1}ZZ }}{(s-M^2_{h_1}) 
    + i \Gamma_{h_1} M_{h_1}} + \frac{g_{h_{2}W_{D}W_{D}} g_{h_{2}ZZ }}{(s-M^2_{h_2}) 
    + i \Gamma_{h_2} M_{h_2}} \right|^{2} 
    \nonumber \\
    &\quad 
    \times \left( 1 + \frac{(s-2M^2_{W_{D}})^2}{8 M^4_{W_{D}}}\right) \left( 1 + \frac{(s-2M^2_{Z})^2}{8 M^4_{Z}}\right)\,.
\end{align}
The vertex is given by
\begin{align}
    g_{h_{1(2)ZZ}} = \frac{v}{2 c^2_w s^2_w} \cos \theta\, (\sin\theta)\,,
\end{align}
where $c^2_{w} = 1 - s^2_{w}$.

\item {\bf $ W_{D}W_{D} \rightarrow f \bar{f }:$}

The cross-section takes
\begin{align}
    \sigma = \frac{1}{16 \pi s} \left( \frac{s-4M^2_{f}}{s-4M^2_{W_D}} \right)^{1/2}\, |M|^2_{ff}
\end{align}
where the amplitude $|M|_{WW}$ is given by
\begin{align}
    |M|^{2}_{ff} &= \frac{4}{9} \left(s - 4 M^2_{f} \right)  \left| \frac{g_{h_{1}W_{D}W_{D}} g_{h_{1}ff }}{(s-M^2_{h_1}) 
    + i \Gamma_{h_1} M_{h_1}} + \frac{g_{h_{2}W_{D}W_{D}} g_{h_{2}ff }}{(s-M^2_{h_2}) 
    + i \Gamma_{h_2} M_{h_2}} \right|^{2} 
    \nonumber \\
    &\quad 
    \times \left( 1 + \frac{(s-2M^2_{W_{D}})^2}{8 M^4_{W_{D}}}\right)\,,
\end{align}
with
\begin{align}
    g_{h_{1(2)ff}} &= - \frac{M_{f}}{v} \cos \theta\, (\sin\theta)\,.
\end{align}

\item {\bf $ W_{D}W_{D} \rightarrow h_{i}h_{j}:$}

The cross-section takes
\begin{align}
    \sigma = \frac{1}{16 \pi s S_{ij}} \left( \frac{s-4M^2_{f}}{s-4M^2_{W_D}} \right)^{1/2}\, |M|^2_{h_{i}h_{j}}
    \,,
\end{align}
where $S_{ij} = 1 (2)$ for $i\neq j (i=j)$ and $i,j = 1,2$. The amplitude $|M|_{h_{i}h_{j}}$ is expressed as
\begin{align}
    |M|^{2}_{h_{i}h_{j}} &=
    \frac{2}{9} \left| \frac{g_{h_{1}W_{D}W_{D}} g_{h_{1}h_{i}h_{j} }}{(s-M^2_{h_1}) 
    + i \Gamma_{h_1} M_{h_1}} + \frac{g_{h_{2}W_{D}W_{D}} g_{h_{2}h_{i}h_{j} }}{(s-M^2_{h_2}) 
    + i \Gamma_{h_2} M_{h_2}} - g_{W_{D}W_{D}h_{i}h_{j}}\right|^{2} 
    \nonumber \\
    &\quad 
    \times \left( 1 + \frac{(s-2M^2_{W_{D}})^2}{8 M^4_{W_{D}}}\right)\,.
\end{align}
The vertices are given by
\begin{align}
    g_{h_{2}h_{2}h_{2}} &=  
    -3 \bigl[ \lambda_{HD} \sin\theta \cos\theta (v_{D} \sin\theta + v \cos\theta) + 2 \lambda_{D}  v_{D} \cos^{3}\theta + 2 \lambda_{H} v \sin^{3}\theta 
    \bigr]
    \,,\nonumber \\
    g_{h_{1}h_{1}h_{1}} &=  
    3 \bigl[ \lambda_{HD} \sin\theta \cos\theta (v_{D} \cos\theta - v \sin\theta) + 2 \lambda_{D} v_{D} \sin^{3}\theta - 2 \lambda_{H} v \cos^{3}\theta 
    \bigr]
    \,,\nonumber \\
    g_{h_{1}h_{2}h_{2}} &=  
    2 (3 \lambda_{D} - \lambda_{HD}) v_{D} \sin\theta \cos^{2}\theta
    + 2 (-3 \lambda_{H} + \lambda_{HD}) v \cos\theta \sin^{2}\theta 
    \nonumber \\
    &\quad
    + \lambda_{HD} (v_{D} \sin^{3}\theta - v \cos^{3}\theta) 
    \,,\nonumber \\
    g_{h_{2}h_{1}h_{1}} &=  
    2 (-3 \lambda_{H} + \lambda_{HD}) v \sin\theta \cos^{2}\theta
    + 2 (-3 \lambda_{D} + \lambda_{HD}) v_{D} \cos\theta \sin^{2}\theta   
    \nonumber \\
    &\quad   
    - \lambda_{HD} (v_{D} \cos^{3}\theta + v \sin^{3}\theta)
    \,,\nonumber \\
    g_{W_{D}W_{D}h_{2}h_{2}} &= 2 \cos^{2}\theta g^2_{D}
    \,,\nonumber\\
    g_{W_{D}W_{D}h_{1}h_{1}} &= 2 \sin^{2}\theta g^2_{D}
    \,,\nonumber\\
    g_{W_{D}W_{D}h_{1}h_{2}} &=  -2\cos\theta \sin\theta g^2_{D} 
    \,.
\end{align}
\end{itemize}    
%%%%%%%%%%%%%%%%%%%%%%%%%%%%%%%%%%%%%%%%%%

%%%%%%%%%%%%%%%%%%%%%%%%%%%%%%%%%%%%%%%%%%
\section{Renormalisation group equations}
\label{apdx:RGEs}
%%%%%%%%%%%%%%%%%%%%%%%%%%%%%%%%%%%%%%%%%%
The beta functions are given by
\begin{align}
    (4\pi)^2\beta_{g_1} &= \frac{81+s_H}{12}g_1^3 \,,\\
    (4\pi)^2\beta_{g_2} &= -\frac{39-s_H}{12}g_2^3 \,,\\
    (4\pi)^2\beta_{g_3} &= -7g_3^3 \,,\\
    (4\pi)^2\beta_{g_D} &= \frac{2n_\psi^2+s_D}{3}g_D^3 \,,\\
    (4\pi)^2\beta_{y_t} &= y_t\left[
    \left(\frac{23}{6}+\frac{2}{3}s_H\right)y_t^2
    -\left(
    8g_3^2 + \frac{17}{12} g_1^2 + \frac{9}{4}g_2^2
    \right)
    \right]
    \,,\\
    (4\pi)^2\beta_{\lambda_H} &=
    6(1+3s_H^2)\lambda_H^2 + \frac{1+s_D^2}{2}\lambda_{HD}^2 - 3g_1^2\lambda_H
    -9g_2^2\lambda_H 
    \nonumber\\
    &\quad
    + \frac{3}{8}g_1^4 + \frac{3}{4}g_1^2g_2^2
    +\frac{9}{8}g_2^4 + 12\lambda_H y_t^2 - 6y_t^4
    \,,\\
    (4\pi)^2\beta_{\lambda_D} &=
    2(1+9s_D^2)\lambda_D^2 + \frac{3+s_H^2}{2}\lambda_{HD}^2
    - 12g_D^2\lambda_D + 6g_D^4
    \,,\\
    (4\pi)^2\beta_{\lambda_{HD}} &=
    6(1+s_H^2)\lambda_H\lambda_{HD} + 2(1+3s_D^2)\lambda_D\lambda_{HD}
    +4s_Hs_D\lambda_{HD}^2
    \nonumber\\
    &\quad
    - \frac{3}{2}g_1^2\lambda_{HD}
    - \frac{9}{2}g_2^2\lambda_{HD} 
    - 6g_D^2\lambda_{HD}
    + 6\lambda_{HD}y_t^2
    \,,
\end{align}
where $s_H$ and $s_D$ are the suppression factors given by
\begin{align}
    s_H = \frac{1+\xi_H h^2/M_{\rm P}^2}{1+(1+6\xi_H)\xi_H  h^2/M_{\rm P}^2}
    \,,\quad
    s_D = \frac{1+\xi_D \phi^2/M_{\rm P}^2}{1+(1+6\xi_D)\xi_D  \phi^2/M_{\rm P}^2}
    \,.
\end{align}
In the absence of the nonminimal coupling, the suppression factor becomes unity.
The beta functions for the nonminimal couplings are as follows:
\begin{align}
    (4\pi)^2\beta_{\xi_H} &=
    \left[
    6(1+s_H)\lambda_H - \frac{3}{2}(g_1^2+3g_2^2) + 6y_t^2
    \right]\left(
    \xi_H + \frac{1}{6}
    \right)
    +(1+s_D)\lambda_{HD}\left(
    \xi_D + \frac{1}{6}
    \right)
    \,,\\
    (4\pi)^2\beta_{\xi_D} &=
    \left[
    2(1+3s_D)\lambda_D - 6g_D^2
    \right]\left(
    \xi_D + \frac{1}{6}
    \right)
    + (3+s_H)\lambda_{HD}\left(
    \xi_H + \frac{1}{6}
    \right)
    \,.
\end{align}
Finally, the anomalous dimensions are
\begin{align}
    (4\pi)^2\gamma_H &=
    -\frac{3}{4}g_1^2 -\frac{9}{4}g_2^2 + 3y_t^2
    \,,\\
    (4\pi)^2\gamma_D &=
    -3g_D^2
    \,.
\end{align}
%%

%\label{appendix}

%\paragraph{Note added.} This is also a good position for notes added
%after the paper has been written.

\bibliographystyle{JHEP}
\bibliography{Inf2CompDM}
%\input{Inf2CompDM.bbl}
%\begin{thebibliography}{99}
%\end{thebibliography}
\end{document}